\documentclass[lettersize,journal]{IEEEtran}
\usepackage{amsmath,amsfonts}
\usepackage{algorithmic}
\usepackage{algorithm}
\usepackage{array}
\usepackage[caption=false,font=normalsize,labelfont=sf,textfont=sf]{subfig}
\usepackage{textcomp}
\usepackage{stfloats}
\usepackage{url}
\usepackage{verbatim}
\usepackage{graphicx}
\usepackage{cite}
\usepackage{cite}
\usepackage{amsmath,amssymb,amsfonts}
\usepackage{algorithmic}
\usepackage{graphicx}
\usepackage{bm}
\usepackage{textcomp}
\usepackage{xcolor}
\usepackage{float}
\usepackage{multirow}
\usepackage{colortbl}
\usepackage{amssymb} 
\usepackage{amsmath}
\usepackage{lettrine}
\usepackage{lipsum}
\begin{document}

\title{Nearest Kronecker Product Decomposition Based Subband Adaptive Filter: Algorithms and Applications}

\author{Jianhong Ye, Haiquan Zhao, \emph{Senior} \emph{Member}, \emph{IEEE}
	\thanks{This work was supported by the National Natural Science Foundation
		of China (NSF) under Grant 62171388, Grant 61871461, and Grant
		61571374. (Corresponding author: Haiquan Zhao).
		
		Jianhong Ye and Haiquan Zhao are with the School of Electrical Engineering,
		Southwest Jiaotong University, Chengdu 611756, China. E-mail address: yjh\_zcl@163.com (J. Ye), hqzhao\_swjtu@126.com (H. Zhao).}}

\maketitle
\begin{abstract}
Recently, the nearest Kronecker product (NKP) decomposition-based normalized least mean square (NLMS-NKP) algorithm has demonstrated superior convergence performance compared to the conventional NLMS algorithm. However, its convergence rate exhibits significant degradation when processing highly correlated input signals. To address this problem, we propose a type-I NKP-based normalized subband adaptive filter (NSAF) algorithm, namely NSAF-NKP-I. Nevertheless, this algorithm incurs substantially higher computational overhead than the NLMS-NKP algorithm. Remarkably, our enhanced type-II NKP-based NSAF (NSAF-NKP-II) algorithm achieves equivalent convergence performance while substantially reducing computational complexity. Furthermore, to enhance robustness against impulsive noise interference, we develop two robust variants: the maximum correntropy criterion-based robust NSAF-NKP (RNSAF-NKP-MCC) and logarithmic criterion-based robust NSAF-NKP (RNSAF-NKP-LC) algorithms. Additionally, detailed analyses of computational complexity, step-size range, and theoretical steady-state performance are provided for the proposed algorithms. To enhance the practicability of the NSAF-NKP-II algorithm in complex nonlinear environments, we further devise two nonlinear implementations: the trigonometric functional link network-based NKP-NSAF (TFLN-NSAF-NKP) and Volterra series expansion-based NKP-NSAF (Volterra-NKP-NSAF) algorithms. In active noise control (ANC) systems, we further propose the filtered-x NSAF-NKP-II (NKP-FxNSAF) algorithm. Simulation experiments in echo cancellation, sparse system identification, nonlinear processing, and ANC scenarios are conducted to validate the superiority of the proposed algorithms over existing state-of-the-art counterparts.
\end{abstract}


\IEEEpeerreviewmaketitle

\section{Introduction}
\lettrine{A}{daptive} filtering algorithms, a quintessential class of online algorithms, are widely used in the design of many real-time systems, such as echo cancellation \cite{guo2021proximal}, speech enhancement \cite{ye2024adaptive}, beamforming \cite{6922488}, and active noise control (ANC) \cite{luo2022hybrid, chen2022distributed, 10890692}. As two of the classical algorithms, the least mean square (LMS) \cite{silva2021skewness} and normalized LMS (NLMS) algorithms \cite{slock1993, kar2016convergence} have garnered extensive attention from researchers due to their simple structure and ease of implementation. However, existing research has shown that the NLMS algorithm exhibits inferior convergence performance when processing highly correlated input signals. To address this issue, researchers have conducted numerous insightful studies in recent decades. As an extension of the NLMS algorithm on the time axis, the affine projection (AP) algorithm \cite{9286540} can achieve faster convergence than the NLMS algorithm in scenarios where the input signal is highly correlated. Nevertheless, the convergence performance of the AP algorithm improves as the projection order increases, which is accompanied by a corresponding rise in computational complexity. Although the recursive least squares (RLS) algorithm \cite{ljung1983theory, towliat2022multi} is derived from the least squares principle and achieves a faster convergence rate relative to the AP algorithm, it incurs greater computational costs due to its inherent inverse matrix operations. To address the need for balancing convergence rate and computational complexity in algorithms for handling strongly correlated input signals, Lee and Gan introduced the normalized subband adaptive filter (NSAF) algorithm \cite{lee2004}. This technology employs analysis filter banks to decompose highly correlated input signals into approximately white subband input signals. Thus, compared with the conventional NLMS algorithm, the NSAF algorithm converges faster while maintaining a comparable level of computational complexity.

In many applications involving the echo environments, the duration of the system impulse response (IR) to be estimated (such as the room impulse response (RIR) or underwater acoustic channel response) is typically very long. To model this response accurately, the length of the adaptive FIR filter must be increased accordingly, which significantly raises the computational complexity of the algorithm. Additionally, a long IR is usually accompanied by a larger eigenvalue spread of the input signal's autocorrelation matrix, leading to a slower convergence rate of the filter. Recently, researchers have proposed an interesting nearest Kronecker product (NKP) decomposition technique \cite{paleologu2018linear, benesty2021, dogariu2022, wang2020time}. This technique transforms the problem of identifying a single long system into that of identifying two shorter systems, thereby effectively reducing the length of the equivalent filter. As the convergence rate of an adaptive algorithm is inversely proportional to the length of the adaptive filter \cite{rupp2015tensor}, the NKP decomposition technique can thus improve the convergence performance of adaptive algorithms, especially for highly sparse systems \cite{bhattacharjee2021, huang2022, wang2021, vadhvana2022, bhattacharjee2020}. So far, this NKP decomposition technique has been successfully exploited in different application frameworks, e.g., system identification \cite{benesty2017}, ANC \cite{ li2025nearest}, nonlinear filtering \cite{bhattacharjee2021}, speech dereverberation \cite{huang2022}, and beamforming \cite{wang2021, vadhvana2022}. In ANC systems, the improved computational efficiency resulting from shorter sub-filter lengths enhances the feasibility of implementing the filtered-x structure on low-power devices. For hearing aid applications, leveraging the NKP structure to model real-time feedback paths can effectively reduce system latency. Specifically, the NKP-based NLMS (NLMS-NKP) algorithm \cite{bhattacharjee2020} improves the convergence rate of the NLMS algorithm for sparse system identification at the expense of a slight increase in computational complexity. In \cite{elisei2019}, the NKP-based RLS (RLS-NKP) algorithm performs slightly worse than the RLS algorithm but achieves a substantial reduction in computational complexity. In addition, the NKP-based AP algorithm has recently been investigated in multichannel ANC scenarios, resulting in the NKP-based multichannel filtered-x AP (NKP-MFxAP) algorithm \cite{li2025nearest}, which demonstrates a more effective suppression of undesired noise compared to the traditional filtered-x AP (FxAP) algorithm. Unfortunately, the decomposed shorter-length adaptive filters in the NKP-MFxAP and RLS-NKP algorithms still involve matrix operations, so their computational complexity is considerably higher than that of the NLMS-NKP algorithm. 

Furthermore, the stable convergence condition of the above-mentioned algorithms depends on the fact that the additive noise follows a Gaussian distribution. However, if the noise is non-Gaussian, (e.g., $\alpha$-stable noise) its impulse response exhibits sharp spikes or occasional bursts \cite{ kumar2022novel}. Under this influence, the performance of the adaptive filter will deteriorate sharply or even diverge. To improve the robustness of the algorithm, methods such as M-estimation \cite{yu2019m}, the least mean $p$-power error (MPE) criterion \cite{shao1993signal, ye2024p, ni2010, 11293772}, and information-theoretic learning-schemes-based correntropy criterion \cite{chen2014steady, huang2017max} have been demonstrated to be feasible robust strategies. In \cite{bhattacharjee2020nearest}, the NKP-based generalized maximum correntropy criterion (NKP-GMCC) and NKP-based generalized hyperbolic secant function (NKP-GHSAF) algorithms achieve stable convergence under impulsive noise scenarios. 

The prerequisite for all the above methods to achieve satisfactory performance is that the adaptive control process is linear. However, complex systems in real-world environments often exhibit strong nonlinear characteristics \cite{zhu2023nonlinear, chen1992neural}. To address this limitation, algorithms based on functional link networks (FLNs) \cite{10806852, zhang2023design}, Volterra series \cite{yu2022interpolated, zhang2024frequency}, kernel methods \cite{shi2023robust, gogineni2022al}, and Spline adaptive filters \cite{7305809, 10535334} have been developed. In \cite{bhattacharjee2021}, Bhattacharjee et al. proposed the trigonometric FLN-based NKP NLMS (TFLN-NKP-NLMS) and Volterra-based NKP NLMS (Volterra-NKP-NLMS) algorithms, which yield favorable performance in the nonlinear system identification tasks. To the best of our knowledge, no studies on subband NKP decomposition algorithms have been reported, either in linear or nonlinear scenarios. Consequently, in light of this reason, we conduct the following research in the context of subband NKP processing.

1) Based on the NKP decomposition technique, the paper proposes a type-I NKP-based NSAF (NSAF-NKP-I) algorithm, which exhibits a faster convergence rate compared to the conventional NSAF algorithm. However, this performance improvement comes at the cost of significant computational overhead. To mitigate this limitation, we further develop a type-II NKP-based NSAF (NSAF-NKP-II) algorithm by modifying the subband NKP model structure. Remarkably, the NSAF-NKP-II algorithm can achieve convergence performance comparable to that of its type-I counterpart while requiring substantially reduced computational resources.

2) To improve the robustness of the NSAF-NKP-II algorithm against impulsive disturbance, the paper proposes two robust variants of the NSAF-NKP-II framework: the maximum correntropy criterion (MCC)-based robust NSAF-NKP-II (RNSAF-NKP-MCC) and the logarithm criterion-based robust NSAF-NKP-II (RNSAF-NKP-LC) algorithms.

3) We establish the theoretical steady-state model and the theoretical step-size range of the NSAF-NKP-II algorithm, and analyze the computational complexity of the proposed NSAF-NKP-I, NSAF-NKP-II, RNSAF-NKP-MCC, and RNSAF-NKP-LC algorithms and compare them with the NSAF, NLMS-NKP, APA-NKP, and RLS-NKP algorithms.

4) To enhance the practicability of the proposed NSAF-NKP-II algorithm in complex nonlinear environments (e.g., asymmetric loudspeaker distortion and symmetrical soft-clipping nonlinearity), while addressing existing gaps in nonlinear subband NKP decomposition methodologies, we {propose two innovative nonlinear subband NKP structures. Specifically, based on the trigonometric FLN (TFLN), we propose the TFLN-based NSAF-NKP-II (TFLN-NKP-NSAF) algorithm. In addition, we develop the Volterra-based NSAF-NKP-II (Volterra-NKP-NSAF) algorithm based on the Volterra series expansion.

5) To verify the effectiveness of the NSAF-NKP-II algorithm in the ANC scenario, we construct a feedforward subband ANC system based on the NKP decomposition technique and develop the filtered-x NSAF-NKP-II (NKP-FxNSAF) algorithm.

\textbf{NOTATION}: $\lvert\lvert .\lvert\lvert_2$ denotes the Euclidean norm, $\lvert\lvert .\lvert\lvert_F$ represents the Frobenius norm, $(.)^{\text{T}}$ is the transpose operation, $\text{E}{\{.\}}$ denotes the expectation operation, $\text{vec}(.)$ denotes the vectorization operation, and $\otimes$ represents the Kronecker product.
\section{ Kronecker Product Decomposition}
\label{sec:guidelines}
Let's consider $\bm{m}_0$ as representing the real-valued impulse response of the unknown system to be identified, which has a size of $D=D_1\times D_2$ with $D_1> D_2$. Next, we can decompose the unknown impulse response $\bm{m}_0$ into $\bm{m}_0=[\bm{t}_1^{\text T}, \bm{t}_2^{\text T},..., \bm{t}_{D_2}^{\text T}]^{\text T}$, where $\{\bm{t}_i\}_{i=1}^{D_2}$ denote the short impulse responses of length $D_1$ and are assumed to be strongly linearly dependent on each other \cite{elisei2019recursive}. According to \cite{paleologu2018linear}, $\bm{m}_0$ can be approximated as $\bm{m}_2\otimes \bm{m}_1=\text{vec}(\bm{m}_1\bm{m}_2^{\text T})$, where $\bm{m}_1$ and $\bm{m}_2$ are two shorter impulse responses of sizes $D_1\times1$ and $D_2\times1$, respectively. To evaluate the approximation accuracy between $\bm{m}_0$ and $\bm{m}_2\otimes \bm{m}_1$, we introduce a normalized misalignment metric defined as 
\begin{equation}
	\begin{split}
		\begin{array}{rcl}
			\begin{aligned}
				\label{001}
				\  \omega\overset{\bigtriangleup}{=}\lvert\lvert\bm{m}_0-\bm{m}_2\otimes \bm{m}_1\lvert\lvert_2/\lvert\lvert\bm{m}_0\lvert\lvert_2.
			\end{aligned}
		\end{array}
	\end{split}
\end{equation}
Next, $\bm{m}_0$ can be reorganized in matrix form as $\bm{M}_0=[\bm{t}_1, \bm{t}_2,..., \bm{t}_{D_2}]$ with dimensions $D_1\times D_2$. Therefore, \eqref{001} can be written as
\begin{equation}
	\begin{split}
		\begin{array}{rcl}
			\begin{aligned}
				\label{002}
				\  \omega\overset{\bigtriangleup}{=}\lvert\lvert\bm{M}_0-\bm{m}_1\bm{m}_2^{\text T}\lvert\lvert_F/\lvert\lvert\bm{M}_0\lvert\lvert_F.
			\end{aligned}
		\end{array}
	\end{split}
\end{equation}
In this form, our goal becomes to find appropriate $\bm{m}_1$ and $\bm{m}_2$ to minimize $\omega$. Interestingly, minimizing the Frobenius norm of  $\omega(\bm{m}_1,\bm{m}_2)$ is equivalent to identifying the nearest rank-1 matrix to $\bm{M}_0$ \cite{van2000}. Singular value decomposition (SVD) provides the mathematical foundation for this optimal approximation \cite{golub2013}. Therefore, we have 
\begin{equation}
	\begin{split}
		\begin{array}{rcl}
			\begin{aligned}
				\label{003}
				\  \bm{M}_0=\bm{ H}_1\bm{\Sigma}\bm{ H}_2^{\text T}=\sum_{i=1}^{D_2}\rho_i\bm{h}_{1,i}\bm{h}_{2,i}^{\text T}.
			\end{aligned}
		\end{array}
	\end{split}
\end{equation}
Here, $\bm\Sigma$ represents an $D_1\times D_2$ rectangular diagonal matrix, containing the singular values of $\bm{M}_0$ sorted in descending order, i.e., $\rho_1\geq\rho_2\geq...\geq\rho_{D_2}\geq0$, the matrices $\bm{ H}_1$ of size $D_1\times D_1$ and $\bm{ H}_2$ of size $D_2\times D_2$ consist of the left and right singular vectors of $\bm{M}_0$, respectively, and the $i$-th columns of $\bm{ H}_1$ and  $\bm{ H}_2$ are $\bm{h}_{1,i}$ and $\bm{h}_{2,i}$, respectively. Using these, the optimal impulse responses of ${\bm m}_1$ and ${\bm m}_2$ are $\breve{\bm m}_1=\sqrt{\rho_1}\bm{h}_{1,1}$ and $\breve{\bm m}_2=\sqrt{\rho_1}\bm{h}_{2,1}$, respectively. However, in the more general case, $\{\bm{t}_i\}_{i=1}^{D_2}$ may not be strongly linearly dependent. In such scenarios, $\bm{m}_0$ can be approximated as
\begin{equation}
	\begin{split}
		\begin{array}{rcl}
			\begin{aligned}
				\label{004}
				\  \bm{m}_0&\approx\sum_{p=1}^P\bm{m}_{2,p}\otimes\bm{m}_{1,p}=\text{vec}\Bigg(\sum_{p=1}^P\bm{m}_{1,p}\bm{m}_{2,p}^{\text T}\Bigg)\\&=\text{vec}\big(\bm{M}_1\bm{M}_2^{\text T}\big)
			\end{aligned}
		\end{array}
	\end{split}
\end{equation}
with $P\leq D_2$, where $\bm{m}_{1,p}$ and $\bm{m}_{2,p}$ stand for the weight vectors of lengths $D_1$ and $D_2$, respectively, and $\bm{M}_1$ of size $D_1\times P$ and $\bm{M}_2$ of size $D_2\times P$ are constituted as follows:
\begin{equation}
	\begin{split}
		\begin{array}{rcl}
			\begin{aligned}
				\label{005}
				\  {\bm M}_1 = \big[{\bm m}_{1,1}\;{\bm m}_{1,2}\;...\;  {\bm m}_{1,P}\big],
			\end{aligned}
		\end{array}
	\end{split}
\end{equation}
\begin{equation}
	\begin{split}
		\begin{array}{rcl}
			\begin{aligned}
				\label{006}
				\   {\bm M}_2 = \big[{\bm m}_{2,1}\;{\bm m}_{2,2}\;...\; {\bm m}_{2,P}\big].
			\end{aligned}
		\end{array}
	\end{split}
\end{equation}

Therefore, the approximation issue of \eqref{002} can be equivalently transformed to minimizing
\begin{equation}
	\begin{split}
		\begin{array}{rcl}
			\begin{aligned}
				\label{007}
				\  \omega\overset{\bigtriangleup}{=}\lvert\lvert\bm{M}_0-\bm{M}_1 \bm{M}_2^{\text T}\lvert\lvert_F/\lvert\lvert\bm{M}_0\lvert\lvert_F,
			\end{aligned}
		\end{array}
	\end{split}
\end{equation}
which leads to the optimal solutions:
\begin{equation}
	\begin{split}
		\begin{array}{rcl}
			\begin{aligned}
				\label{008}
				\  \breve{\bm M}_1 =& \big[\breve{\bm m}_{1,1}\;\breve{\bm m}_{1,2}\;...\;  \breve{\bm m}_{1,P}\big]\\&=\big[\sqrt{\rho_1}\bm{h}_{1,1}, \sqrt{\rho_2}\bm{h}_{1,2},...,\sqrt{\rho_P}\bm{h}_{1,P}\big],
			\end{aligned}
		\end{array}
	\end{split}
\end{equation}
\begin{equation}
	\begin{split}
		\begin{array}{rcl}
			\begin{aligned}
				\label{009}
				\  \breve{\bm M}_2 =& \big[\breve{\bm m}_{2,1}\;\breve{\bm m}_{2,2}\;...\;  \breve{\bm m}_{2,P}\big]\\&=\big[\sqrt{\rho_1}\bm{h}_{2,1}, \sqrt{\rho_2}\bm{h}_{2,2},...,\sqrt{\rho_P}\bm{h}_{2,P}\big].
			\end{aligned}
		\end{array}
	\end{split}
\end{equation}

Based on \eqref{008} and \eqref{009}, the optimal approximation of $\bm{m}_0$ is
\begin{equation}
	\begin{split}
		\begin{array}{rcl}
			\begin{aligned}
				\label{010}
				\  \breve{\bm m}_0 =\sum_{p=1}^P\breve{\bm m}_{2,p}\otimes\breve{\bm m}_{1,p} =\sum_{p=1}^P\rho_p{\bm h}_{2,p}\otimes{\bm h}_{1,p}.
			\end{aligned}
		\end{array}
	\end{split}
\end{equation}
\section{Proposed NSAF-NKP algorithms}
\subsection{Derivation of NSAF-NKP-I algorithm}
In linear system identification scenarios, the input vector $\bm{x}_r=[x_r,x_{r-1},...,x_{r-D+1}]^{\text T}$ and the desired signal $d_r$ have the following relationship at time instant $r$:
\begin{equation}
	\begin{split}
		\begin{array}{rcl}
			\begin{aligned}
				\label{011}
				\ d_r=\bm{x}^{\text T}_r\bm{m}_0+v_r,
			\end{aligned}
		\end{array}
	\end{split}
\end{equation}
where $v_r$ denotes the additive noise, which is statistically independent of the input vector $\bm{x}_r$.

Fig. 1 describes the structure of the NSAF-NKP-I algorithm with $N$ subbands. The "synthesizer" functions to combine the short sub-filters $\hat{\bm m}_{r,2}$ and $\hat{\bm m}_{r,1}$ into the final adaptive filter $\hat{\bm m}_{r}$. By resorting to the analysis filters whose impulse responses are represented as $\{\bm{f}_j\}_{j=1}^N$ of length $L$, the signal ${d}_r$ can be decomposed into the subband desired signal ${d}_{r,j}$. In addition, we introduce the desired vector $\hat{\bm{d}}_{r}$ of size $1 \times L $ as 
\begin{align}
	\label{013} 
	\hat{\bm{d}}_{r}\overset{\bigtriangleup}{=}[{d}_r, {d}_{r-1},...,{d}_{r-L+1}].
\end{align}

We refer to the set formed by subband desired signals $\{{d}_{r,j}\}_{j=1}^N$ as the subband desired vector $\bm{d}_r$, i.e.,
\begin{align}
	\label{015} 
	\bm{d}_r\overset{\bigtriangleup}{=}[{d}_{r,1}, {d}_{r,2},..., {d}_{r,N}]=\hat{\bm{d}}_{r}\textbf{F},
\end{align}
where $\textbf{F}\overset{\bigtriangleup}{=}[\bm{f}_1,\bm{f}_2,...,\bm{f}_N]$ denotes the analysis filter matrix with a size of $L\times N$. 
\begin{figure}[htb]
	\centering
	\includegraphics[scale=0.47] {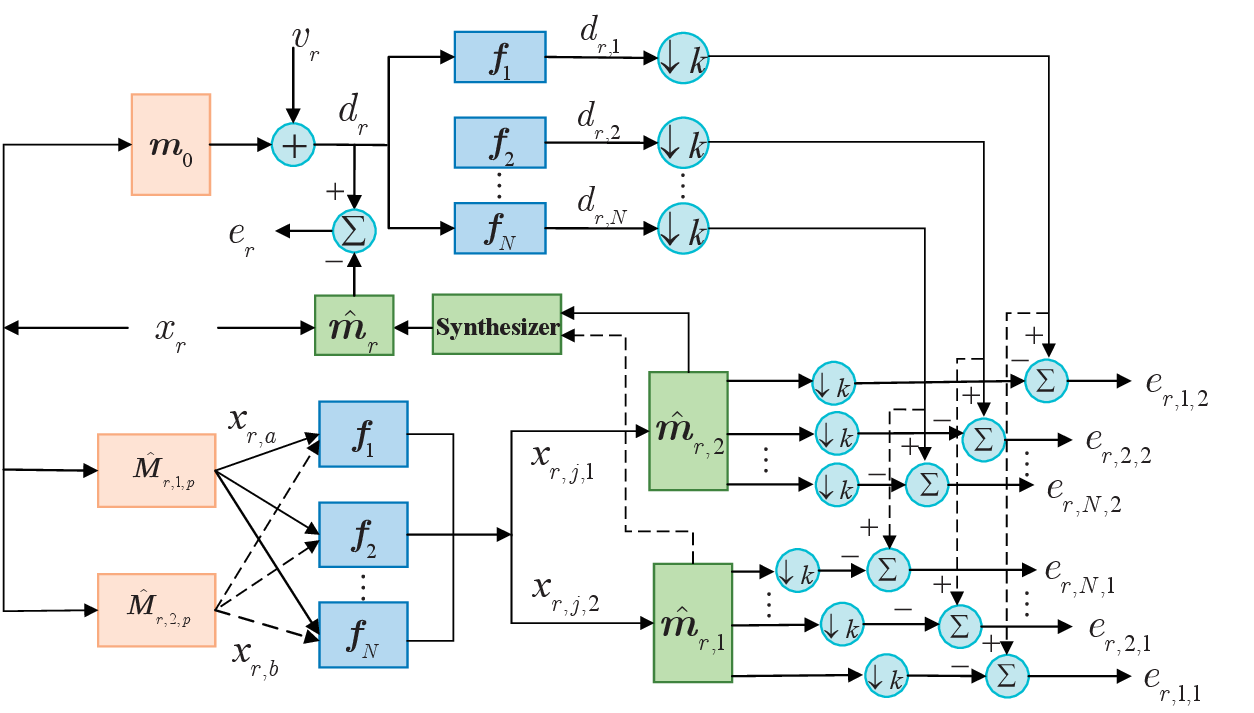}
	\vspace{-1em} \caption{Structure of NSAF-NKP-I.}
	\label{Fig0}
\end{figure}

Using a similar decomposition method to \eqref{010}, the adaptive filter $\hat{\bm m}_r$ can be decomposed as
\begin{align}
	\label{016} 
\hat{\bm m}_r=\sum_{p=1}^P\hat{\bm m}_{r,2,p}\otimes\hat{\bm m}_{r,1,p}, 
\end{align}
where $\hat{\bm m}_{r,1,p} \in \mathbb{R}^{D_1\times 1}$ and $\hat{\bm m}_{r,2,p} \in \mathbb{R}^{D_2\times 1}$ represent the $p$-th column of sub-filter matrices $\widetilde{\bm M}_{r,1}\overset{\bigtriangleup}{=}[\hat{\bm m}_{r,1,1},\hat{\bm m}_{r,1,2},...,\hat{\bm m}_{r,1,P}]\in \mathbb{R}^{D_1\times P}$ and $\widetilde{\bm M}_{r,2}\overset{\bigtriangleup}{=}[\hat{\bm m}_{r,2,1},\hat{\bm m}_{r,2,2},...,\hat{\bm m}_{r,2,P}]\in \mathbb{R}^{D_2\times P }$, respectively. Furthermore, we define the following relationship \cite{harville1998matrix}:
\begin{equation}
	\begin{split}
		\begin{array}{rcl}
			\begin{aligned}
				\label{017}
				\ \hat{\bm m}_{r,2,p}\otimes\hat{\bm m}_{r,1,p}&\overset{\bigtriangleup}{=}\big[\hat{\bm m}_{r,2,p}\otimes\bm{ I}_{D_1}\big]\hat{\bm m}_{r,1,p}\\&\overset{\bigtriangleup}{=}\big[\bm{I}_{D_2}\otimes\hat{\bm m}_{r,1,p}\big]\hat{\bm m}_{r,2,p},
			\end{aligned}
		\end{array}
	\end{split}
\end{equation}
where $\bm{ I}_{D_1}$ and $\bm{ I}_{D_2}$ are identity matrices with sizes of $D_1\times D_1$ and $D_2\times D_2$, respectively.

By inserting \eqref{017} into \eqref{016}, we obtain
\begin{equation}
	\begin{split}
		\begin{array}{rcl}
			\begin{aligned}
				\label{018}
				\ \hat{\bm m}_r&=\sum_{p=1}^P\hat{\bm M}_{r,2,p}\hat{\bm m}_{r,1,p}=\sum_{p=1}^P\hat{\bm M}_{r,1,p}\hat{\bm m}_{r,2,p},
			\end{aligned}
		\end{array}
	\end{split}
\end{equation}
where $\hat{\bm M}_{r,1,p}\overset{\bigtriangleup}{=}\big[\bm{I}_{D_2}\otimes\hat{\bm m}_{r,1,p}\big]$ and $\hat{\bm M}_{r,2,p}\overset{\bigtriangleup}{=}\big[\hat{\bm m}_{r,2,p}\otimes\bm{ I}_{D_1}\big]$ are rectangular matrices of sizes $D_1D_2\times D_2$ and $D_1D_2\times D_1$, respectively.

The error signal $e_r$ is calculated by
\begin{equation}
	\begin{split}
		\begin{array}{rcl}
			\begin{aligned}
				\label{019}
				\ e_r=d_r-\hat{\bm m}^{\text T}_r\bm{x}_r.
			\end{aligned}
		\end{array}
	\end{split}
\end{equation}

By taking \eqref{018} into \eqref{019}, we obtain the error signals of the sub-filters as
\begin{equation}
	\begin{split}
		\begin{array}{rcl}
			\begin{aligned}
				\label{033}
				\ e_r&=e_{r,1}=d_r-\sum_{p=1}^P\hat{\bm m}_{r,1,p}^{\text T}\hat{\bm M}_{r,2,p}^{\text T}\bm{x}_r\\&=d_r-\sum_{p=1}^P\hat{\bm m}_{r,1,p}^{\text T}\bm{x}_{r,2,p}=d_r-\hat{\bm m}_{r,1}^{\text T}\bm{x}_{r,b},
			\end{aligned}
		\end{array}
	\end{split}
\end{equation}
\begin{equation}
	\begin{split}
		\begin{array}{rcl}
			\begin{aligned}
				\label{034}
				\ e_r&=e_{r,2}=d_r-\sum_{p=1}^P\hat{\bm m}_{r,2,p}^{\text T}\hat{\bm M}_{r,1,p}^{\text T}\bm{x}_r\\&=d_r-\sum_{p=1}^P\hat{\bm m}_{r,2,p}^{\text T}\bm{x}_{r,1,p}=d_r-\hat{\bm m}_{r,2}^{\text T}\bm{x}_{r,a},
			\end{aligned}
		\end{array}
	\end{split}
\end{equation}
where
\begin{equation}
	\begin{split}
		\begin{array}{rcl}
			\begin{aligned}
				\label{035}
				\ \bm{x}_{r,2,p}=\hat{\bm M}_{r,2,p}^{\text T}\bm{x}_r,
			\end{aligned}
		\end{array}
	\end{split}
\end{equation}
\begin{equation}
	\begin{split}
		\begin{array}{rcl}
			\begin{aligned}
				\label{036}
				\ \bm{x}_{r,b}=\big[\bm{x}_{r,2,1}^{\text T},\bm{x}_{r,2,2}^{\text T},...,\bm{x}_{r,2,P}^{\text T}\big]^{\text T},
			\end{aligned}
		\end{array}
	\end{split}
\end{equation}
\begin{equation}
	\begin{split}
		\begin{array}{rcl}
			\begin{aligned}
				\label{025_a}
				\ \hat{\bm m}_{r,1}=[\hat{\bm m}_{r,1,1}^{\text T},\hat{\bm m}_{r,1,2}^{\text T},...,\hat{\bm m}_{r,1,P}^{\text T} ]^{\text T},
			\end{aligned}
		\end{array}
	\end{split}
\end{equation}
\begin{equation}
	\begin{split}
		\begin{array}{rcl}
			\begin{aligned}
				\label{037}
				\ \bm{x}_{r,1,p}=\hat{\bm M}_{r,1,p}^{\text T}\bm{x}_r,
			\end{aligned}
		\end{array}
	\end{split}
\end{equation}
\begin{equation}
	\begin{split}
		\begin{array}{rcl}
			\begin{aligned}
				\label{038}
				\ \bm{x}_{r,a}=\big[\bm{x}_{r,1,1}^{\text T},\bm{x}_{r,1,2}^{\text T},...,\bm{x}_{r,1,P}^{\text T}\big]^{\text T},
			\end{aligned}
		\end{array}
	\end{split}
\end{equation}
\begin{equation}
	\begin{split}
		\begin{array}{rcl}
			\begin{aligned}
				\label{027_a}
				\ \hat{\bm m}_{r,2}=[\hat{\bm m}_{r,2,1}^{\text T},\hat{\bm m}_{r,2,2}^{\text T},...,\hat{\bm m}_{r,2,P}^{\text T} ]^{\text T}.
			\end{aligned}
		\end{array}
	\end{split}
\end{equation}

\textbf{Remark 1}. The initial values of $\hat{\bm m}_{r,1,p}$ and $\hat{\bm m}_{r,2,p}$ are performed as $\hat{\bm m}_{0,1,p}=[\lambda\;\textbf{0}_{D_1-1}^{\text T}]^{\text T}$ and $\hat{\bm m}_{0,2,p}=[\lambda\;\textbf{0}_{D_2-1}^{\text T}]^{\text T}$ with $0<\lambda \leq 1$, respectively. To the best of our knowledge, all the existing adaptive filtering algorithms based on NKP decomposition adopt the aforementioned initial value setting method (i.e., the original method). Furthermore, we proposed a novel method for setting the initial values, i.e.,
\[
\hat{\bm M}_{0,2}=\begin{pmatrix}
	\lambda & 0  & \cdots & 0 \\
	0 & \lambda  & \cdots & 0 \\
	\vdots  & \vdots & \ddots & \vdots\\
	0 & 0  & \cdots & \lambda \\
\end{pmatrix}_{D_2 \times P}.
\]
We refer to this initial value setting method as the YIM method, and its effectiveness has been verified in Fig. 8(e). However, this problem is not the main focus of this paper. To be consistent with other NKP adaptive filtering algorithms, we still choose to use the original method.

Then, we difine the input rectangular matrices $\bm{X}_{r,a}$ of size $PD_2\times L$ and $\bm{X}_{r,b}$ of size $PD_1\times L$ as
\begin{align}
	\label{039} 
	\bm{X}_{r,a}\overset{\bigtriangleup}{=}[\bm{x}_{r,a},\bm{x}_{r-1,a},...,\bm{x}_{r-L+1,a}],
\end{align}
\begin{align}
	\label{040} 
	\bm{X}_{r,b}\overset{\bigtriangleup}{=}[\bm{x}_{r,b},\bm{x}_{r-1,b},...,\bm{x}_{r-L+1,b}].
\end{align}

By utilizing the analysis filters $\{\bm{f}_j\}_{j=1}^N$, the subband input vectors $\bm{x}_{r,j,1}\overset{\bigtriangleup}{=}\big[\bm{x}_{r,j,1,1}^{\text T},\bm{x}_{r,j,1,2}^{\text T},...,\bm{x}_{r,j,1,P}^{\text T}\big]^{\text T}$ and $\bm{x}_{r,j,2}\overset{\bigtriangleup}{=}\big[\bm{x}_{r,j,2,1}^{\text T},\bm{x}_{r,j,2,2}^{\text T},...,\bm{x}_{r,j,2,P}^{\text T}\big]^{\text T}$ of the sub-filters $\hat{\bm m}_{r,2}$ and $\hat{\bm m}_{r,1}$ can be collected into the subband input rectangular matrices $\bar{\bm X}_{r,a}$ and $\bar{\bm X}_{r,b}$, respectively, i.e.,
\begin{align}
	\label{041} 
	\bar{\bm X}_{r,a}\overset{\bigtriangleup}{=}[\bm{x}_{r,1,1},\bm{x}_{r,2,1},...,\bm{x}_{r,N,1}]=\bm{X}_{r,a}\textbf{F},
\end{align}
\begin{align}
	\label{042} 
	\bar{\bm X}_{r,b}\overset{\bigtriangleup}{=}[\bm{x}_{r,1,2},\bm{x}_{r,2,2},...,\bm{x}_{r,N,2}]=\bm{X}_{r,b}\textbf{F},
\end{align}
where the sizes of $\bar{\bm X}_{r,a}$ and $\bar{\bm X}_{r,b}$ are $PD_2\times N$ and $PD_1\times N$, respectively.

Considering the Kronecker product decomposition method, the subband forms of the error signals in \eqref{033} and \eqref{034} can be written as
\begin{equation}
	\begin{split}
		\begin{array}{rcl}
			\begin{aligned}
				\label{032_b}
				\ e_{r,j}=d_{r,j}-\hat{\bm m}_{r,1}^{\text T}\bm{x}_{r,j,2}=d_{r,j}-\hat{\bm m}_{r,2}^{\text T}\bm{x}_{r,j,1}.
			\end{aligned}
		\end{array}
	\end{split}
\end{equation}

To establish the weight recursion ways of the adaptive filters $\hat{\bm m}_{r,1}$ and $\hat{\bm m}_{r,2}$ in the subband domain, the cost function can be expressed as
\begin{equation}
	\begin{split}
		\begin{array}{rcl}
			\begin{aligned}
				\label{029_a}
				\ J_{\hat{\bm m}_{r,2}}\big(\hat{\bm m}_{r,1}\big)=\frac{1}{2}\sum_{j=1}^N\frac{ e_{r,j}^2}{\lvert\lvert \bm{x}_{r,j,2}\lvert\lvert_2^2},
			\end{aligned}
		\end{array}
	\end{split}
\end{equation}
\begin{equation}
	\begin{split}
		\begin{array}{rcl}
			\begin{aligned}
				\label{030_a}
				\ J_{\hat{\bm m}_{r,1}}\big(\hat{\bm m}_{r,2}\big)=\frac{1}{2}\sum_{j=1}^N\frac{ e_{r,j}^2}{\lvert\lvert \bm{x}_{r,j,1}\lvert\lvert_2^2}.
			\end{aligned}
		\end{array}
	\end{split}
\end{equation}

By resorting the stochastic gradient descent rule, the weight update for the sub-filters $\hat{\bm m}_{r,1}$ and $\hat{\bm m}_{r,2}$ can be written as
\begin{equation}
	\begin{split}
		\begin{array}{rcl}
			\begin{aligned}
				\label{031_a}
				\ \hat{\bm m}_{r+k,1} = \hat{\bm m}_{r,1}+\mu_1\sum_{j=1}^{N}\frac{\bm{x}_{r,j,2}e_{r,j}}{\lvert\lvert \bm{x}_{r,j,2}\lvert\lvert_2^2+\delta},
			\end{aligned}
		\end{array}
	\end{split}
\end{equation}
\begin{equation}
	\begin{split}
		\begin{array}{rcl}
			\begin{aligned}
				\label{032_a}
				\ \hat{\bm m}_{r+k,2} = \hat{\bm m}_{r,2}+\mu_2\sum_{j=1}^{N}\frac{\bm{x}_{r,j,1}e_{r,j}}{\lvert\lvert \bm{x}_{r,j,1}\lvert\lvert_2^2+\delta},
			\end{aligned}
		\end{array}
	\end{split}
\end{equation}
where $\mu_1>0$ and $\mu_2>0$ are the step-size, $k$ is a positive integer that determines the interval between iterations, $\delta$ is a small positive number to avoid division by zero.

So far, \eqref{016}, \eqref{041}, \eqref{042}, \eqref{032_b}, \eqref{031_a}, and \eqref{032_a} constitute the proposed NSAF-NKP-I algorithm. 
\subsection{Derivation of NSAF-NKP-II algorithm}
\begin{figure}[htb]
	\centering
	\includegraphics[scale=0.5] {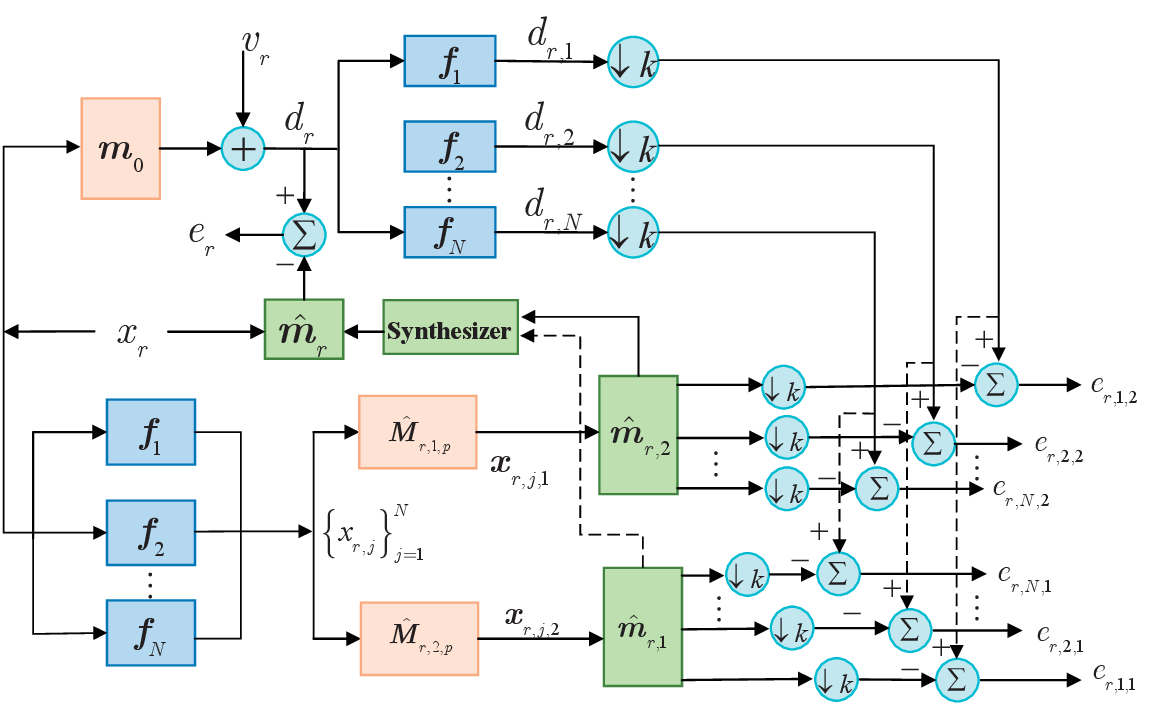}
	\vspace{-1em} \caption{Structure of NSAF-NKP-II.}
	\label{Fig01}
\end{figure}
By resorting to the analysis filters, the input vector $\bm{x}_r$ can be decomposed into the subband input vector $\{\bm{x}_{r,j}\}_{j=1}^N$. Then, we introduce the input rectangular matrix ${\bm{X}}_{r}$ of size $D\times L$ as
\begin{align}
	\label{012a} 
	{\bm{X}}_{r}\overset{\bigtriangleup}{=}[\bm{x}_r,\bm{x}_{r-1},...,\bm{x}_{r-L+1}].
\end{align}

The subband input rectangular matrix $\bar{\bm X}_r \in \mathbb{R}^{D\times N}$ is obtained by collecting the subband input vectors $\{\bm{x}_{r,j}\}_{j=1}^N$, i.e.,
\begin{align}
	\label{014a} 
	\bar{\bm X}_r\overset{\bigtriangleup}{=}[\bm{x}_{r,1},\bm{x}_{r,2},...,\bm{x}_{r,N}]={\bm{X}}_{r}\textbf{F}.
\end{align}

Fig. 2 describes the structure of the NSAF-NKP-II algorithm. Similar to the derivation of the NSAF-NKP-I algorithm, the $p$-th subband input vectors $\bm{x}_{r,j,2,p}$ and $\bm{x}_{r,j,1,p}$ for sub-filters $\hat{\bm m}_{r,1}$ and $\hat{\bm m}_{r,2}$ can be calculated by
\begin{equation}
	\begin{split}
		\begin{array}{rcl}
			\begin{aligned}
				\label{023}
				\ \bm{x}_{r,j,2,p}=\hat{\bm M}_{r,2,p}^{\text T}\bm{x}_{r,j},
			\end{aligned}
		\end{array}
	\end{split}
\end{equation}
\begin{equation}
	\begin{split}
		\begin{array}{rcl}
			\begin{aligned}
				\label{026}
				\ \bm{x}_{r,j,1,p}=\hat{\bm M}_{r,1,p}^{\text T}\bm{x}_{r,j}.
			\end{aligned}
		\end{array}
	\end{split}
\end{equation}

Equations \eqref{016}, \eqref{032_b}, \eqref{031_a}, \eqref{032_a}, \eqref{023}, and \eqref{026} constitute the proposed NSAF-NKP-II algorithm. 

\textbf{Remark 2.} It is crucial to emphasize that the NSAF-NKP-I algorithm and the NSAF-NKP-II algorithm differ solely in their model architectures. Analysis and simulation demonstrate that this structural discrepancy does not impact algorithmic performance; however, as shown in Fig. 4 and Table II of Section IV, the computational complexity of the NSAF-NKP-II algorithm is significantly lower than that of the NSAF-NKP-I algorithm.
\subsection{Performance analysis}
To establish the steady-state mean square deviation (MSD) model of the NKP-NSAF-II algorithm, we define the subband error signal of the filter $\hat{\bm m}_r$ as
\begin{equation}
	\begin{split}
		\begin{array}{rcl}
			\begin{aligned}
				\label{A1}
				\ e_{r,j} = \bm{x}_{r,j}^{\text T}\widetilde{\bm m}_r + v_{r,j} = \zeta_{r,j} +  v_{r,j},
			\end{aligned}
		\end{array}
	\end{split}
\end{equation}
where $\widetilde{\bm m}_r\overset{\bigtriangleup}{=}\bm{m}_0-\hat{\bm m}_r $ donotes the weight error vector, $v_{r,j}$ is the subband noise vector, and $\zeta_{r,j}$ represents an excess error. Similar to (10), the long unknown system $\bm{m}_0$ can be expressed as
\begin{equation}
	\begin{split}
		\begin{array}{rcl}
			\begin{aligned}
				\label{A2}
				\bm{m}_0= \sum_{p=1}^{P} \bm{m}_{2,p}^o\otimes\bm{m}_{1,p}^o.
			\end{aligned}
		\end{array}
	\end{split}
\end{equation}

Then, let us define the decomposed weight error vectors as
\begin{equation}
	\begin{split}
		\begin{array}{rcl}
			\begin{aligned}
				\label{A3}
\ \widetilde{\bm m}_{r,2,p}={\bm m}_{2,p}^o - \hat{\bm m}_{r,2,p},
			\end{aligned}
		\end{array}
	\end{split}
\end{equation}
\begin{equation}
	\begin{split}
		\begin{array}{rcl}
			\begin{aligned}
				\label{A4}
				\ \widetilde{\bm m}_{r,1,p}={\bm m}_{1,p}^o - \hat{\bm m}_{r,1,p}.
			\end{aligned}
		\end{array}
	\end{split}
\end{equation}

Based on (14) and \eqref{A2}, the excess error signal can be approximated as
\begin{equation}
	\begin{split}
		\begin{array}{rcl}
			\begin{aligned}
				\label{A5}
				\ \zeta_{r,j} &= \bm{x}_{r,j}^{\text T}\Big[\sum_{p=1}^{P} \bm{m}_{2,p}^o\otimes\bm{m}_{1,p}^o-\sum_{p=1}^P \hat{\bm m}_{r,2,p} \otimes \hat{\bm m}_{r,1,p}\Big]\\&\approx \bm{x}_{r,j}^{\text T}\Big[\sum_{p=1}^{P} \widetilde{\bm m}_{r,2,p}\otimes\hat{\bm{m}}_{r,1,p}+\sum_{p=1}^{P} \hat{\bm m}_{r,2,p}\otimes\widetilde{\bm{m}}_{r,1,p}\Big]\\&=\zeta_{r,j,1}+\zeta_{r,j,2}.
			\end{aligned}
		\end{array}
	\end{split}
\end{equation}
In \eqref{A5}, we make an approximation by omitting term $\bm{x}_{r,j}^{\text T}\sum_{p=1}^{P} \widetilde{\bm m}_{r,2,p}\otimes\widetilde{\bm{m}}_{r,1,p}$, as it comprises the higher-order terms of the weight error vector that become insignificant under steady-state stage. In addition, we collect $\widetilde{\bm m}_{r,2}\overset{\bigtriangleup}{=}[\widetilde{\bm m}_{r,2,1}^{\text T},...,\widetilde{\bm m}_{r,2,P}^{\text T}]^{\text T}$ and $\widetilde{\bm m}_{r,1}\overset{\bigtriangleup}{=}[\widetilde{\bm m}_{r,1,1}^{\text T},...,\widetilde{\bm m}_{r,1,P}^{\text T}]^{\text T}$. Based on these, the terms $\zeta_{r,j,1}$ and $\zeta_{r,j,2}$ can be reexpressed as
\begin{equation}
	\begin{split}
		\begin{array}{rcl}
			\begin{aligned}
				\label{A6}
				\ \zeta_{r,j,1}=\sum_{p=1}^{P}\bm{x}_{r,j}^{\text T} \Big[\widetilde{\bm m}_{r,2,p}\otimes\hat{\bm{m}}_{r,1,p}\Big]=\bm{x}_{r,j,1}^{\text T}\widetilde{\bm m}_{r,2},
			\end{aligned}
		\end{array}
	\end{split}
\end{equation}
\begin{equation}
	\begin{split}
		\begin{array}{rcl}
			\begin{aligned}
				\label{A7}
				\ \zeta_{r,j,2}=\sum_{p=1}^{P}\bm{x}_{r,j}^{\text T} \Big[\hat{\bm m}_{r,2,p}\otimes\widetilde{\bm{m}}_{r,1,p}\Big]=\bm{x}_{r,j,2}^{\text T}\widetilde{\bm m}_{r,1}.
			\end{aligned}
		\end{array}
	\end{split}
\end{equation}

To facilitate the following derivation, some assumptions are made as follows:

{\it Assumption 1}: The weight error vectors $\widetilde{\bm m}_{r,1}$ and $\widetilde{\bm m}_{r,2}$ are mutually independent, and both are uncorrelated with the input signal.

{\it Assumption 2}: The subband error signals $e_{r,j,1}$ and $e_{r,j,2}$ of sub-filters are asymptotically uncorrelated with the input vectors $\bm{x}_{r,j,2}$ and $\bm{x}_{r,j,1}$, respectively. The zero-mean additive noise $v_r$ is uncorrelated of other signals. This assumption is commonly utilized in the performance analysis of the adaptive filtering algorithms.

{\it Assumption 3}: The excess error signals $\zeta_{r,j,1}$ and $\zeta_{r,j,2}$ are zero-mean and mutually independent when the input signal is zero-mean white, which is utilized to eliminate the cross terms $\text{E}[\zeta_{r,j,1}\zeta_{r,j,2}]$.

Based on \eqref{A3} and \eqref{A4}, we further obtain
\begin{equation}
	\begin{split}
		\begin{array}{rcl}
			\begin{aligned}
				\label{A9}
				\ \widetilde{\bm m}_{r+k,1} = \widetilde{\bm m}_{r,1}-\mu_1\sum_{j=1}^{N}\frac{\bm{x}_{r,j,2}e_{r,j}}{\lvert\lvert \bm{x}_{r,j,2}\lvert\lvert_2^2},
			\end{aligned}
		\end{array}
	\end{split}
\end{equation}
\begin{equation}
	\begin{split}
		\begin{array}{rcl}
			\begin{aligned}
				\label{A10}
				\ \widetilde{\bm m}_{r+k,2} = \widetilde{\bm m}_{r,2}-\mu_2\sum_{j=1}^{N}\frac{\bm{x}_{r,j,1}e_{r,j}}{\lvert\lvert \bm{x}_{r,j,1}\lvert\lvert_2^2}.
			\end{aligned}
		\end{array}
	\end{split}
\end{equation}
Then, by left-multiplying both sides of \eqref{A9} and \eqref{A10} by their respective transposes and taking the expectation, we get
\begin{equation}
	\begin{split}
		\begin{array}{rcl}
			\begin{aligned}
				\label{A11}
				\ &\text{E}\big\{\lvert\lvert\widetilde{\bm m}_{r+k,1}\lvert\lvert_2^2\big\}=\text{E}\big\{\lvert\lvert\widetilde{\bm m}_{r,1}\lvert\lvert_2^2\big\}\\&-2\mu_1\sum_{j=1}^N\text{E}\Big\{\frac{e_{r,j}\zeta_{r,j,2}}{\lvert\lvert \bm{x}_{r,j,2}\lvert\lvert_2^2}\Big\}+\mu_1^2\sum_{j=1}^N\text{E}\Big\{\frac{e_{r,j}^2}{\lvert\lvert \bm{x}_{r,j,2}\lvert\lvert_2^2}\Big\},
			\end{aligned}
		\end{array}
	\end{split}
\end{equation}
\begin{equation}
	\begin{split}
		\begin{array}{rcl}
			\begin{aligned}
				\label{A12}
				\ &\text{E}\big\{\lvert\lvert\widetilde{\bm m}_{r+k,2}\lvert\lvert_2^2\big\}=\text{E}\big\{\lvert\lvert\widetilde{\bm m}_{r,2}\lvert\lvert_2^2\big\}\\&-2\mu_2\sum_{j=1}^N\text{E}\Big\{\frac{e_{r,j}\zeta_{r,j,1}}{\lvert\lvert \bm{x}_{r,j,1}\lvert\lvert_2^2}\Big\}+\mu_2^2\sum_{j=1}^N\text{E}\Big\{\frac{e_{r,j}^2}{\lvert\lvert \bm{x}_{r,j,1}\lvert\lvert_2^2}\Big\}.
			\end{aligned}
		\end{array}
	\end{split}
\end{equation}

Due to the subband signals are uncorrelated and the fluctuation in $\lvert\lvert \bm{x}_{r,j,1}\lvert\lvert^2$ and $\lvert\lvert \bm{x}_{r,j,2}\lvert\lvert^2$ from one iteration to the next can be assumed to be small enough \cite{4176554}. When the sub-filters converge to the steady-state stage, i.e., $r \rightarrow \infty$, the energy conservation relations \eqref{A11} and \eqref{A12} can be approximated as
\begin{equation}
	\begin{split}
		\begin{array}{rcl}
			\begin{aligned}
				\label{A13}
				\ 2\text{E}\big\{e_{r,j}\zeta_{r,j,2}\big\} = \mu_1\text{E}\big\{e_{r,j}^2\big\},
			\end{aligned}
		\end{array}
	\end{split}
\end{equation}
\begin{equation}
	\begin{split}
		\begin{array}{rcl}
			\begin{aligned}
				\label{A14}
				\ 2\text{E}\big\{e_{r,j}\zeta_{r,j,1}\big\} = \mu_2\text{E}\big\{e_{r,j}^2\big\}.
			\end{aligned}
		\end{array}
	\end{split}
\end{equation}

Based on assumptions 2 and 3, by putting \eqref{A1} and \eqref{A5} in to \eqref{A11} and \eqref{A12}, we obtain
\begin{equation}
	\begin{split}
		\begin{array}{rcl}
			\begin{aligned}
				\label{A15}
				\ (2-\mu_1)\text{E}\big\{\zeta_{r,j,2}^2\big\} = \mu_1\big(\text{E}\big\{\zeta_{r,j,1}^2\big\}+\text{E}\big\{v_{r,j}^2\big\}\big),
			\end{aligned}
		\end{array}
	\end{split}
\end{equation}
\begin{equation}
	\begin{split}
		\begin{array}{rcl}
			\begin{aligned}
				\label{A16}
				\ (2-\mu_2)\text{E}\big\{\zeta_{r,j,1}^2\big\} = \mu_2\big(\text{E}\big\{\zeta_{r,j,2}^2\big\}+\text{E}\big\{v_{r,j}^2\big\}\big).
			\end{aligned}
		\end{array}
	\end{split}
\end{equation}

Combining \eqref{A15} and \eqref{A16}, we obtain
\begin{equation}
	\begin{split}
		\begin{array}{rcl}
			\begin{aligned}
				\label{A17}
				\ \text{E}\big\{\zeta_{r,j,1}^2\big\}=\frac{\mu_2\text{E}\big\{v_{r,j}^2\big\}}{2-\mu_1-\mu_2},
			\end{aligned}
		\end{array}
	\end{split}
\end{equation}
\begin{equation}
	\begin{split}
		\begin{array}{rcl}
			\begin{aligned}
				\label{A18}
				\ \text{E}\big\{\zeta_{r,j,2}^2\big\}=\frac{\mu_1\text{E}\big\{v_{r,j}^2\big\}}{2-\mu_1-\mu_2}.
			\end{aligned}
		\end{array}
	\end{split}
\end{equation}

Since $\sigma_{v_r}^2=\frac{1}{N}\sum_{j=1}^N\sigma_{v_{r,j}}^2$, the steady-state excess mean-square error (EMSE) of the NSAF-NKP-II algorithm can be defined as
\begin{equation}
	\begin{split}
		\begin{array}{rcl}
			\begin{aligned}
				\label{A20}
				\ \text{EMSE} &\overset{\bigtriangleup}{=}\frac{1}{N}\sum_{j=1}^N\text{E}\{\zeta_{r,j}^2\}= \frac{1}{N}\sum_{j=1}^N\Big[\text{E}\{\zeta_{r,j,1}^2\}+\text{E}\{\zeta_{r,j,2}^2\}\Big]\\&=\frac{(\mu_1+\mu_2)\text{E}\big\{v_{r}^2\big\}}{2-\mu_1-\mu_2}.
			\end{aligned}
		\end{array}
	\end{split}
\end{equation}

Therefore, to ensure the stability of the NSAF-NKP-II algorithm, the denominators in \eqref{A17} and \eqref{A18} should be greater than zero, i.e.,
\begin{equation}
	\begin{split}
		\begin{array}{rcl}
			\begin{aligned}
				\label{A21}
				0<\mu_1+\mu_2<2.
			\end{aligned}
		\end{array}
	\end{split}
\end{equation}
If we set the same step-size, i.e., $\mu_1=\mu_2$, we obtain
\begin{equation}
	\begin{split}
		\begin{array}{rcl}
			\begin{aligned}
				\label{A22}
				0<\mu<1,
			\end{aligned}
		\end{array}
	\end{split}
\end{equation}
whose verification can be observed in Fig. 6 (b). From the results, this step-size range is smaller than that of the NSAF algorithm, which is $0<\mu<2$ \cite{lee2004}.
\section{Robust variants of NSAF-NKP-II}
To enhance the robustness of the NSAF-NKP-II algorithm to impulsive noise, we introduce the scaling functions $\mathcal{X}_{r,j,1}$ and $\mathcal{X}_{r,j,2}$, respectively, for subband $j$ to generalize \eqref{031_a} and \eqref{032_a} as
\begin{equation}
	\begin{split}
		\begin{array}{rcl}
			\begin{aligned}
				\label{053}
				\ \hat{\bm m}_{r+k,1} = \hat{\bm m}_{r,1}+\mu_1\sum_{j=1}^{N}\mathcal{X}_{r,j,1}\frac{\bm{x}_{r,j,2}e_{r,j,1}}{\lvert\lvert \bm{x}_{r,j,2}\lvert\lvert_2^2+\delta},
			\end{aligned}
		\end{array}
	\end{split}
\end{equation}
\begin{equation}
	\begin{split}
		\begin{array}{rcl}
			\begin{aligned}
				\label{054}
				\ \hat{\bm m}_{r+k,2} = \hat{\bm m}_{r,2}+\mu_2\sum_{j=1}^{N}\mathcal{X}_{r,j,2}\frac{\bm{x}_{r,j,1}e_{r,j,2}}{\lvert\lvert \bm{x}_{r,j,1}\lvert\lvert_2^2+\delta}.
			\end{aligned}
		\end{array}
	\end{split}
\end{equation}

\textbf{Remark 3.} The condition for \eqref{053} and \eqref{054} to achieve stable convergence under impulsive noise is that the value of $\mathcal{X}_{r,j,1}$ and $\mathcal{X}_{r,j,2}$ are small when impulsive noise occurs. To achieve this effect, we design the RNSAF-NKP-MCC and RNSAF-NKP-LC algorithms by maximizing correntropy-based and minimizing logarithm-based cost functions, respectively. 

\subsection{Derivation of RNSAF-NKP-MCC algorithm}
In information theory learning, the correntropy is an effective tool to measure the local similarity of two arbitrary random variables $\mathcal{R}$ and $\mathcal{Y}$ \cite{chen2014steady}, which is defined as
\begin{align}
	\label{043} 
	V(\mathcal{R},\mathcal{Y})=\text{E}\big\{\kappa(\mathcal{R},\mathcal{Y})\big\},
\end{align}
where $\kappa(.)$ represents the shift-invariant Mercer kernel. The Gaussian function is often used as a kernel function for correntropy because of its many good properties such as smooth and symmetric. Therefore, we have
\begin{align}
	\label{044} 
	\kappa(\mathcal{R},\mathcal{Y})=\frac{1}{\sqrt{2\pi}\sigma}\text{exp}\Big(-\frac{(\mathcal{R}-\mathcal{Y})^2}{2\sigma^2}\Big),
\end{align}
where $\sigma>0$ denotes the kernel bandwidth. 

Then, we establish the robust cost functions based on the NSAF-NKP-II framework as follows:
\begin{align}
	\label{045} 
	J_{{\hat{\bm m}_{r,1,\text{MCC}}}}=\frac{1}{2\psi}\sum_{j=1}^N\text{exp}\Big(-\psi\frac{\Big[d_{r,j}-\hat{\bm m}_{r,1}^{\text T}\bm{x}_{r,j,2}\Big]^2}{\lvert\lvert\bm{x}_{r,j,2}\lvert\lvert_2^2}\Big),
\end{align}
\begin{align}
	\label{046} 
	J_{{\hat{\bm m}_{r,2,\text{MCC}}}}=\frac{1}{2\psi}\sum_{j=1}^N\text{exp}\Big(-\psi\frac{\Big[d_{r,j}-\hat{\bm m}_{r,2}^{\text T}\bm{x}_{r,j,1}\Big]^2}{\lvert\lvert\bm{x}_{r,j,1}\lvert\lvert_2^2}\Big),
\end{align}
where $\psi=\frac{1}{2\sigma^2}$ represents the kernel parameter.

By combining with the gradient ascent method and taking the partial derivatives of $\hat{\bm m}_{r,1}$ and $\hat{\bm m}_{r,2}$ on both sides of \eqref{045} and \eqref{046}, respectively, and then setting the results to zero, the proposed RNSAF-NKP-MCC algorithm is established as
\begin{equation}
	\begin{split}
		\begin{array}{rcl}
			\begin{aligned}
				\label{047}
				\ &\hat{\bm m}_{r+k,1} = \hat{\bm m}_{r,1}\\&+\mu_1\sum_{j=1}^{N}\frac{\bm{x}_{r,j,2}e_{r,j,1}}{\lvert\lvert \bm{x}_{r,j,2}\lvert\lvert_2^2+\delta}\times\underbrace{\text{exp}\Big(-\psi\frac{e_{r,j,1}^2}{\lvert\lvert \bm{x}_{r,j,2}\lvert\lvert_2^2}\Big)}_{(a)},
			\end{aligned}
		\end{array}
	\end{split}
\end{equation}
\begin{equation}
	\begin{split}
		\begin{array}{rcl}
			\begin{aligned}
				\label{048}
				\ &\hat{\bm m}_{r+k,2} = \hat{\bm m}_{r,2}\\&+\mu_2\sum_{j=1}^{N}\frac{\bm{x}_{r,j,1}e_{r,j,2}}{\lvert\lvert \bm{x}_{r,j,1}\lvert\lvert_2^2+\delta}\times\underbrace{\text{exp}\Big(-\psi\frac{e_{r,j,2}^2}{\lvert\lvert \bm{x}_{r,j,1}\lvert\lvert_2^2}\Big)}_{(b)}.
			\end{aligned}
		\end{array}
	\end{split}
\end{equation}
Clearly, the error signals $\{e_{r,j,1}, e_{r,j,2}\}$ exhibit considerably elevated magnitudes during impulsive noise occurrences. However, the terms $(a)$ and $(b)$ in \eqref{047} and \eqref{048} will become very small to ensure that a smaller step-size is selected, which has been proven to effectively avoid the adverse effects of outliers in impulse noise on the update of filter weights \cite{song2013nor}.
\begin{figure}[htp]
	\centering  
	\includegraphics[scale=0.35] {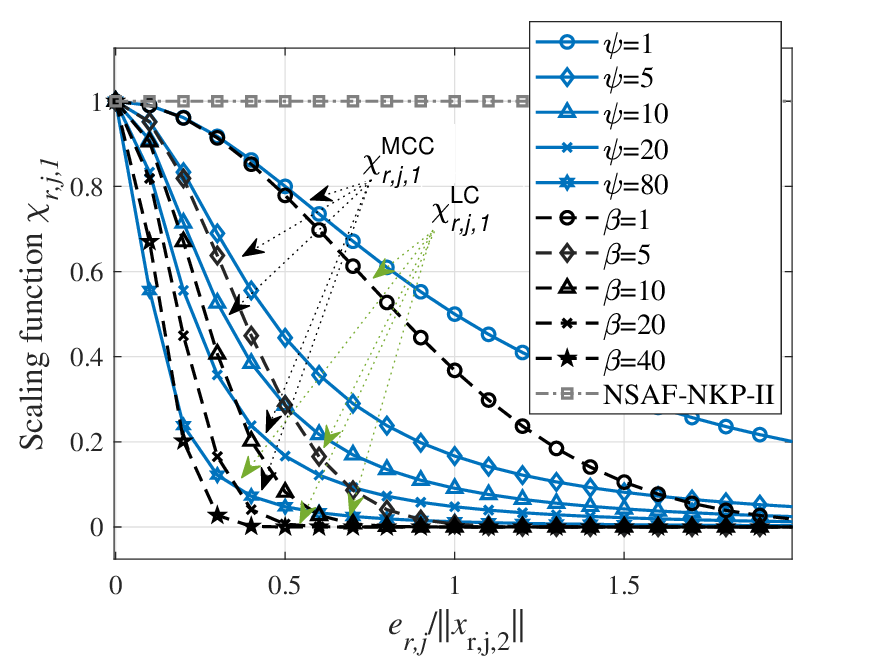}  
	\caption{Scale function of the $i$th subband.}
	\label{Fig16}
\end{figure}
\subsection{Derivation of RNSAF-NKP-LC algorithm}
Inspired by \cite{sayin2014novel}, we extend the logarithmic-based cost function to the NSAF-NKP-II structure to establish
\begin{align}
	\label{049} 
	J_{{\hat{\bm m}_{r,1,\text{LC}}}}=\frac{1}{2\beta}\sum_{j=1}^N\text{ln}\Big(1+\beta\frac{\Big[d_{r,j}-\hat{\bm m}_{r,1}^{\text T}\bm{x}_{r,j,2}\Big]^2}{\lvert\lvert\bm{x}_{r,j,2}\lvert\lvert_2^2}\Big),
\end{align}
\begin{align}
	\label{050} 
	J_{{\hat{\bm m}_{r,2,\text{LC}}}}=\frac{1}{2\beta}\sum_{j=1}^N\text{ln}\Big(1+\beta\frac{\Big[d_{r,j}-\hat{\bm m}_{r,2}^{\text T}\bm{x}_{r,j,1}\Big]^2}{\lvert\lvert\bm{x}_{r,j,1}\lvert\lvert_2^2}\Big),
\end{align}
where $\beta>0$ is a adjustable parameter. 

Taking the gradient of $J_{{\hat{\bm m}_{r,1,\text{LC}}}}$ and $J_{{\hat{\bm m}_{r,2,\text{LC}}}}$ with respect to $\hat{\bm m}_{r,1}$ and $\hat{\bm m}_{r,2}$, respectively, the RNSAF-NKP-LC algorithm is established as
\begin{equation}
	\begin{split}
		\begin{array}{rcl}
			\begin{aligned}
				\label{051}
				\ &\hat{\bm m}_{r+k,1} = \hat{\bm m}_{r,1}+\sum_{j=1}^{N}\frac{\mu_1\frac{\bm{x}_{r,j,2}e_{r,j,1}}{\lvert\lvert \bm{x}_{r,j,2}\lvert\lvert_2^2+\delta}}{1+\beta\big(e_{r,j,1}/\lvert\lvert \bm{x}_{r,j,2}\lvert\lvert_2\big)^2},
			\end{aligned}
		\end{array}
	\end{split}
\end{equation}
\begin{equation}
	\begin{split}
		\begin{array}{rcl}
			\begin{aligned}
				\label{052}
				\ &\hat{\bm m}_{r+k,2} = \hat{\bm m}_{r,2}+\sum_{j=1}^{N}\frac{\mu_2\frac{\bm{x}_{r,j,1}e_{r,j,2}}{\lvert\lvert \bm{x}_{r,j,1}\lvert\lvert_2^2+\delta}}{1+\beta\big(e_{r,j,2}/\lvert\lvert \bm{x}_{r,j,1}\lvert\lvert_2\big)^2}.
			\end{aligned}
		\end{array}
	\end{split}
\end{equation}

\textbf{Remark 4}. Fig. 3 shows the shapes of the scaling functions $\chi_{r,j,1}^{\text {MCC}}$ and $\chi_{r,j,1}^{\text {LC}}$ with respect to the term $e_{r,j,1}/\lvert\lvert \bm{x}_{r,j,2}\lvert\lvert$ utilizing different $\psi$ and $\beta$. Clearly, whenever the impulsive interference happens, $e_{r,j,1}/\lvert\lvert \bm{x}_{r,j,2}\lvert\lvert$ exhibits a large amplitude. Subsequently, the scaling functions $\chi_{r,j,1}^{\text {MCC}}$ and $\chi_{r,j,1}^{\text {LC}}$ will limit the step-size to a smaller value to ensure that the outliers do not have adverse effects on the weight update process of the RNSAF-NKP-MCC and RNSAF-NKP-LC algorithms. Noteworthy, a small step-size can mitigate the negative impact of outliers on the update of filter weights \cite{song2013normalized}. Of course, if the impulsive interference does not occur, the value of $e_{r,j,1}/\lvert\lvert \bm{x}_{r,j,2}\lvert\lvert$ is very small, and thus the convergence behavior of the RNSAF-NKP-MCC and RNSAF-NKP-LC algorithms is approximately similar to that of the NSAF-NKP-II algorithm due to $\chi_{r,j,1}^{\text {MCC}}$ and $\chi_{r,j,1}^{\text {LC}}$ approaching 1. For example, as shown in Fig. 9, the performance of the RNSAF-NKP-MCC and RNSAF-NKP-LC algorithms under Gaussian noise scenario in terms of the convergence rate and steady-state NMSD are not damaged by the introduction of corresponding robustness criteria compared with the NSAF-NKP-II algorithm.
\begin{table*}[htp]
	\renewcommand\arraystretch{1.1}
	\tabcolsep = 1cm
	
	\setlength{\abovecaptionskip}{0cm}
	\setlength{\belowcaptionskip}{0cm}
	\begin{center}
		\caption{Computational complexity of different algorithms for every $k$ input samples.}
		\begin{tabular}{c c c c c c c} 
			\hline
			\rowcolor{gray!30}	 Algorithms &$\times/\div$   \\			
			\hline
		NLMS 	 &$k(3D+2)$	\\
			\rowcolor{gray!30}NSAF &$3ND+2N+(D+1)LN$\\
			NLMS-NKP &$k(5PD+3PD_1+3PD_2+4)$ \\
			\rowcolor{gray!30}RLS-NKP&$k(5PD+5P^2D_1^2+5P^2D_2^2+3PD_1+3PD_2)$ \\
			APA-NKP&$k[5PD+(2C+1+C^2)(PD_1+PD_2)+2C^2]$ \\
			\rowcolor{gray!30}NSAF-NKP-II &$PD+4PND+3NPD_2+3NPD_1+4N+(D+1)LN$ \\
			NSAF-NKP-I &$PD+4DPL+(PD_1+1)LN+(PD_2+1)LN+3PND_1+3PND_2+4N$\\
			\rowcolor{gray!30}RNSAF-NKP-MCC &$PD+4PND+3NPD_2+3NPD_1+4N+(D+1)LN+NP(D_1+D_2)+8N$ \\
		RNSAF-NKP-LC &$PD+4PND+3NPD_2+3NPD_1+4N+(D+1)LN+NP(D_1+D_2)+8N$ \\
			\hline
		\end{tabular}
	\end{center}
\end{table*}
\begin{table*}[htp]
	\renewcommand\arraystretch{1.5}
	\tabcolsep = 0.05cm
	\setlength{\abovecaptionskip}{0cm}
	\setlength{\belowcaptionskip}{0cm}
	\begin{center}
		\caption{Comparison of time comsumption of different algorithms.}
		\begin{tabular}{c c c c c c c c c c c c} 
			\hline
			\rowcolor{gray!30}	Algorithms&NLMS&NSAF&NLMS-NKP&RLS-NKP&APA-NKP (C=5)&APA-NKP (C=15)&NSAF-NKP-I&NSAF-NKP-II &RNSAF-NKP-LC&RNSAF-NKP-MCC  \\			
			\hline
			Runtime 	 &0.1872s&0.3121s&5.04s&6.8008s&18.2783s &51.4564s&90.2347s&\bf{5.8983s}&6.3114s&6.3824s	\\
			\hline
		\end{tabular}
	\end{center}
\end{table*}
\subsection{Complexity of Algorithm}
\begin{figure}[htp]
	\centering  
	\includegraphics[scale=0.29] {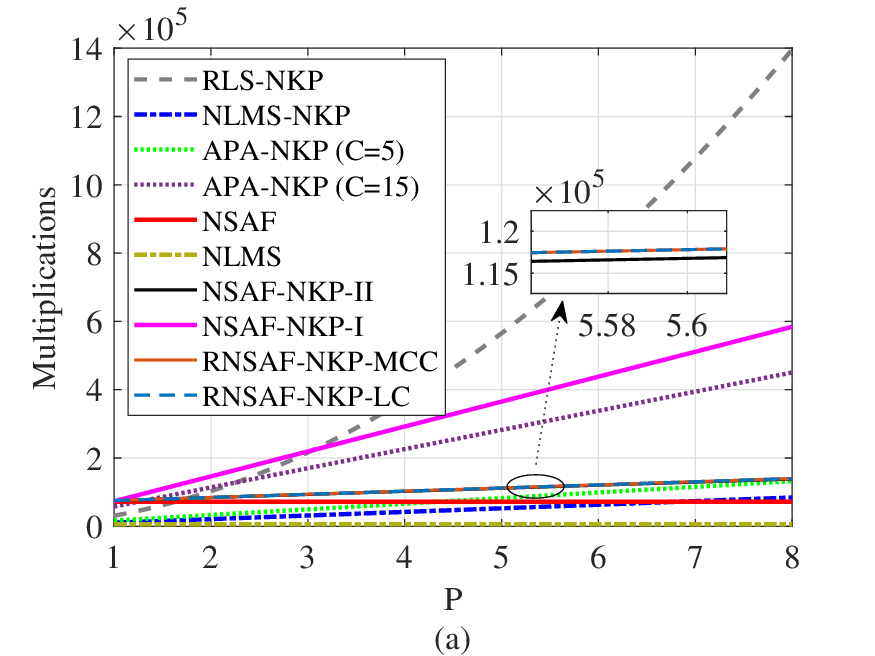}  
	\hspace{0.001ex}									 
	\includegraphics[scale=0.29] {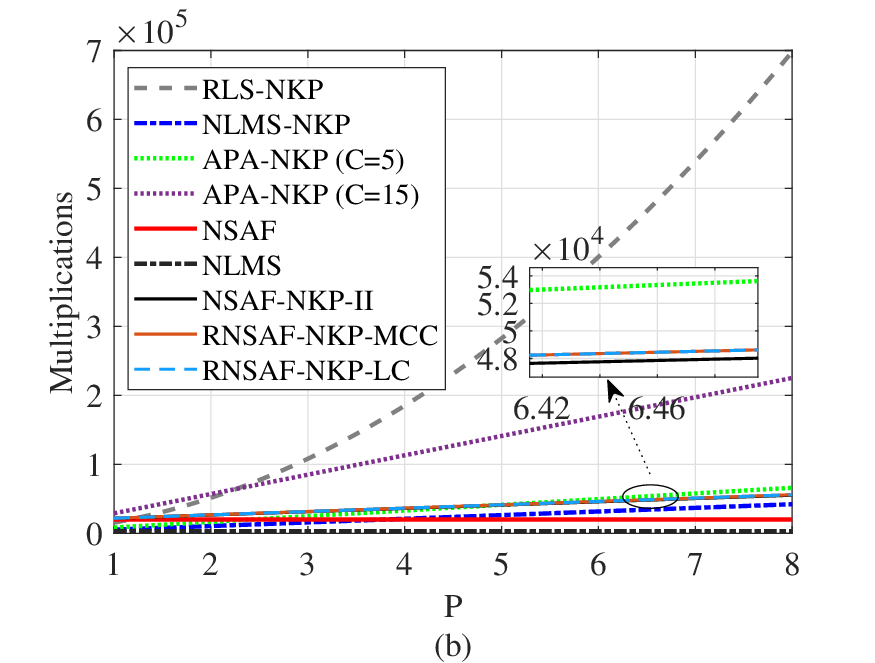}								  
	\caption{The total multiplications per $k$ input samples of the NLMS, NSAF, NLMS-NKP, RLS-NKP, APA-NKP, NSAF-NKP-I, NSAF-NKP-II, RNSAF-NKP-MCC, and RNSAF-NKP-LC algorithms. (a) [$L=33$, $N=4$, $D=500$, $D_1=25$, $D_2=20$, and $k=4$] for the parameter setting; (b) [$L=17$, $N=2$, and $k=2$] for the parameter setting, and the other parameters of the algorithms are the same as those in Fig. 3(a). }
	\label{Fig5}
\end{figure}
Table I summarizes the computational complexity of the NLMS \cite{slock1993}, NSAF \cite{lee2004}, NLMS-NKP \cite{bhattacharjee2020}, RLS-NKP \cite{elisei2019}, NKP-based AP (APA-NKP) \cite{li2025nearest}, NSAF-NKP-I, NSAF-NKP-II, RNSAF-NKP-MCC, and RNSAF-NKP-LC algorithms in the context of system identification, which count the quantities of multiplication operations for every $k$ input samples. Note that, the NSAF structure has inherent computational complexity for partitioning both the extended input vector $\bm{x}_r$ and the desired output signal $d_r$, which needs $LN(D$+$1)$ multiplications. The projection order $C$ governs the convergence and steady-state performance of the APA-NKP algorithm, when $C=1$, the APA-NKP algorithm reduces to the NLMS-NKP algorithm.  Typically, a higher projection order ensures better performance but results in higher computational complexity. Fig. 4(a) plots the total multiplications per $k$ input samples of the algorithms with $N=4$. The NLMS-NKP algorithm requires a greater number of multiplications compared to the NLMS algorithm. As we will see in the simulation below, the proposed NSAF-NKP-I algorithm attains a faster convergence rate than the NLMS-NKP algorithm, albeit with an increased multiplication burden. The proposed NSAF-NKP-II algorithm matches the convergence performance of the NSAF-NKP-I while significantly reducing computational complexity. Note that the computational complexity of the RLS-NKP and APA-NKP ($C=15$) algorithms are much greater than that of the proposed algorithms in most cases. Furthermore, the proposed RNSAF-NKP-MCC and RNSAF-NKP-LC algorithms achieve robustness to impulse noise at the cost of only a minimal increase in computational overhead compared to the NSAF-NKP-II algorithm. Fig. 4(b) shows the total multiplications per $k$ input samples of the algorithms with $N=2$. We can also draw some conclusions similar to those in Fig. 4(a). 

Additionally, we conducted rigorous numerical simulations measuring practical execution times in the Table II. All experiments were implemented in MATLAB on a laptop equipped with an AMD Ryzen 9 4000HS with Radeon Graphics 3.0 GHz processor and 16 GB of RAM. The input signal consisted of 30000 samples. The runtime in Table II was obtained by summing the time required for 50 Monte Carlo experiments and then taking the average. Clearly, the runtime of the NSAF-NKP-II algorithm is slightly slower than that of the NLMS-NKP algorithm because the former requires the use of an analysis filter to decompose the input signal and the desired signal into subband signals. The runtime of the NSAF-NKP-II algorithm is slightly faster than that of the RLS-NKP algorithm.  At the same time, it is only 32.2\% of the runtime of the APA-NKP ($C=5$) algorithm, and only 6.5\% of the runtime of the NSAF-NKP-I algorithm, because the APA-NKP algorithm involves matrix multiplication and inversion. In addition, the runtimes of the proposed RNSAF-NKP-LC and RNSAF-NKP-MCC algorithms are approximately 0.4131s and 0.4841s slower than that of the NSAF-NKP-II algorithm respectively, achieving robustness against impulsive noise. 
\section{NKP-based Nonlinear subband algorithms}
Nonlinear interference is ubiquitous in practical application scenarios. To improve the robustness of the NSAF-NKP-II algorithm against such interference, we propose two robust variants of the NSAF-NKP-II algorithm: the TFLN-NKP-NSAF and Volterra-NKP-NSAF algorithms. These robust algorithms employ functional expansion blocks (FEBs) to project the input vector $\bm{x}_r=[x_r,x_{r-1},...,x_{r-B+1}]^{\text{T}}$ into a higher-dimensional feature space through nonlinear transformations. Fig. 5 describes the implementation process of the nonlinear subband NKP algorithms. Crucially, the distinction between TFLN-NKP-NSAF and Volterra-NKP-NSAF lies solely in their distinct nonlinear projection mechanisms, which lead to different nonlinear modeling capabilities. For instance, the FEB module is designed by leveraging trigonometric functional expansion, through which the resultant high-dimensional input vector $\bm{g}_r$ can be mathematically formulated as:
\begin{equation}
	\begin{split}
		\begin{array}{rcl}
			\begin{aligned}
				\label{055}
				\ &\bm{g}_r=\Big[\bm{x}^{\text T}_r, \text{sin}(\pi x_r),\text{cos}(\pi x_r),...,\\&\text{cos}(A\pi x_r),\text{sin}(\pi x_{r-1}),\text{cos}(\pi x_{r-1})\\&,...,\text{cos}(A\pi x_{r-1}),...,\text{cos}(A\pi x_{r-B+1})\Big]^{\text T},
			\end{aligned}
		\end{array}
	\end{split}
\end{equation}
where $A$ denotes the expansion order and the length of $\bm{g}_r$ is $D=(2A+1)B$. Similarly, the FEB module's design incorporates the Volterra series expansion, thereby we obtain
\begin{equation}
	\begin{split}
		\begin{array}{rcl}
			\begin{aligned}
				\label{056}
				\ &\bm{g}_r=\big[\bm{x}^{\text T}_r, x^2_r,x^2_{r-1},...,x^2_{r-B+1},\\&x_rx_{r-1},...,x_rx_{r-B+1},x_{r-1}x_{r-2},...,\\&x_{r-1}x_{r-B+1},...,x_{r-B+2}x_{r-B+1}\big]^{\text T},
			\end{aligned}
		\end{array}
	\end{split}
\end{equation}
with the length of $D=(B^2+3B)/2$.
\begin{table*}[htp]
	\renewcommand\arraystretch{1}
	\tabcolsep = 0.5cm
	\setlength{\abovecaptionskip}{0cm}
	\setlength{\belowcaptionskip}{0cm}
	\begin{center}
		\caption{Comparison of proposed and existing methods.}
		\begin{tabular}{c c c c c c c} 
			\hline
			\rowcolor{gray!30}	 Algorithms &Domain&Correlated&Matrix Operation& Noise&Nonlinear&Reference \\	
			\rowcolor{gray!30}	  &&input&Inverse Operation& Robustness&& \\			
			\hline
			NLMS-NKP 	 &Fullband &Moderate&$\times$&Gaussian only&$\times$&\cite{bhattacharjee2020}	\\
			RLS-NKP &Fullband&Strong&$\checkmark$&Gaussian only&$\times$&\cite{elisei2019}\\
			APA-NKP &Fullband&Strong&$\checkmark$&Gaussian only&$\times$&\cite{li2025nearest} \\
			NKP-GMCC&Fullband&Moderate&$\times$&non-Gaussian&$\times$&\cite{bhattacharjee2020nearest}	 \\
			NKP-GHSAF&Fullband&Moderate&$\times$&non-Gaussian&$\times$&	\cite{bhattacharjee2020nearest} \\
			TFLN-NKP-NLMS &Fullband&Moderate&$\times$&Gaussian only&$\checkmark$&\cite{bhattacharjee2021}	 \\
			Volterra-NKP-NLMS &Fullband&Moderate&$\times$&Gaussian only&$\checkmark$&\cite{bhattacharjee2021}	\\
			\rowcolor{gray!30}NSAF-NKP-I &Subband&Strong&$\times$&Gaussian only&$\times$&Proposed	 \\
			\rowcolor{gray!30}NSAF-NKP-II &Subband&Strong&$\times$&Gaussian only&$\times$&Proposed	 \\
			\rowcolor{gray!30}RNSAF-NKP-LC &Subband&Strong&$\times$&non-Gaussian&$\times$&Proposed	 \\
			\rowcolor{gray!30}RNSAF-NKP-MCC &Subband&Strong&$\times$&non-Gaussian&$\times$&Proposed	 \\
			\rowcolor{gray!30}Volterra-NKP-NSAF &Subband&Strong&$\times$&Gaussian only&$\checkmark$&Proposed	 \\
			\rowcolor{gray!30}TFLN-NKP-NSAF &Subband&Strong&$\times$&Gaussian only&$\checkmark$&Proposed	 \\
			\hline
		\end{tabular}
	\end{center}
\end{table*}
\begin{table}[htp]
	\renewcommand\arraystretch{1 }
	\scriptsize
	\centering
	\caption{Pseudocode of the RNSAF-NKP-LC and RNSAF-NKP-MCC algorithms}
	\label{table_3}
	\begin{tabular}{lc}
		\hline
		\text{Initialization:} $\hat{\bm m}_{0,1,p}=[\lambda\;0\;...\;0]^{\text T},\;p=1,2,...,P$\\
		\;\;\;\;\;\;\;\;\;\;\;\;\;\;\;\;\;\;\;\;\;$\hat{\bm m}_{0,2,p}=[\lambda\;0\;...\;0]^{\text T},\;p=1,2,...,P$\\ 
		\hline
		{\bf for} each time instant $r$ {\bf do}:\\
		\;\;\;/ * \textsl{Trigonometric\;function\;expansion}\;\;\;\;\;\;\;\;\;\;\;* /\\
		\;\;\;$\bm{g}_r=\big[\bm{x}^{\text T}_r, \text{sin}(\pi x_r),\text{cos}(\pi x_r),...,\text{cos}(A\pi x_r),\text{sin}(\pi x_{r-1})$\\
		\;\;\;\;\;\;\;\;\;\;\;\;\;\;\;\;$,\text{cos}(\pi x_{r-1}),...,\text{cos}(A\pi x_{r-1}),...,$\\
		\;\;\;\;\;\;\;\;\;\;\;\;\;\;\;\;$\text{cos}(A\pi x_{r-B+1})\big]^{\text T}$\\
		\;\;\;/ * \textsl{Volterra\;series\;expansion}\;\;\;\;\;\;\;\;\;\;\;\;\;\;\;\;\;\;\;\;\;\;\;\;* /\\
		\;\;\;$\bm{g}_r=\big[\bm{x}^{\text T}_r, x^2_r,x^2_{r-1},...,x^2_{r-B+1},x_rx_{r-1},...,$\\
		\;\;\;\;\;\;\;\;\;\;\;\;\;\;\;\;$x_rx_{r-B+1},x_{r-1}x_{r-2},...,x_{r-1}x_{r-B+1}$\\
		\;\;\;\;\;\;\;\;\;\;\;\;\;\;\;\;$,...,x_{r-B+2}x_{r-B+2}\big]^{\text T}$\\ 
		\;\;\;$\hat{\bm{X}}_{r}=[\bm{g}_r, \bm{g}_{r-1},..., \bm{g}_{r-L+1}]$\\
		\;\;\;$\hat{\bm{d}}_{r}=[{d}_r, {d}_{r-1},..., {d}_{r-L+1}]$\\
		\;\;\;\;\;\;\;\;{\bf if} $\text{mod}(r,k)==0$\\
		\;\;\;\;\;\;\;\;\;\;\;\;\;\;\;$\bm{X}_r=[\bm{x}_{r,1},\bm{x}_{r,2},...,\bm{x}_{r,N}]=\hat{\bm{X}}_{r}\textbf{F}$\\
		\;\;\;\;\;\;\;\;\;\;\;\;\;\;\;$\bm{d}_r=[{d}_{r,1},{d}_{r,2},...,{d}_{r,N}]=\hat{\bm{d}}_{r}\textbf{F}$\\
		\;\;\;\;\;\;\;\;\;\;\;\;\;\;\;\;\;\;\;\;\;\;\text{{\bf for} each subband} $j$\\
		\;\;\;\;\;\;\;\;\;\;\;\;\;\;\;\;\;\;\;\;\;\;\;\;\;\;\;\;$\bm{x}_{r,j,2,p}=\hat{\bm M}_{r,2,p}^{\text T}\bm{x}_{r,j}$\\
		\;\;\;\;\;\;\;\;\;\;\;\;\;\;\;\;\;\;\;\;\;\;\;\;\;\;\;\;$\bm{x}_{r,j,2}=\big[\bm{x}_{r,j,2,1}^{\text T},\bm{x}_{r,j,2,2}^{\text T},...,\bm{x}_{r,j,2,P}^{\text T}\big]^{\text T}$\\
		\;\;\;\;\;\;\;\;\;\;\;\;\;\;\;\;\;\;\;\;\;\;\;\;\;\;\;\;$e_{r,j,1}=d_{r,j}-\hat{\bm m}_{r,1}^{\text T}\bm{x}_{r,j,2}$\\
		\;\;\;\;\;\;\;\;\;\;\;\;\;\;\;\;\;\;\;\;\;\;\bf{end}\\
		\;\;\;\;\;\;\;\;\;\;\;\;\;\;\;\;\;\;\;\;\;\;$\hat{\bm m}_{r+k,1}=\hat{\bm m}_{r,1}+\mu_1\sum_{j=1}^N\frac{\bm{x}_{r,j,2}e_{r,j,1}}{\lvert\lvert\bm{x}_{r,j,2}\lvert\lvert_2^2+\delta}$\\
		\;\;\;\;\;\;\;\;\;\;\;\;\;\;\;\;\;\;\;\;\;\;\text{{\bf for} each subband} $j$\\
		\;\;\;\;\;\;\;\;\;\;\;\;\;\;\;\;\;\;\;\;\;\;\;\;\;\;\;\;$\bm{x}_{r,j,1,p}=\hat{\bm M}_{r,1,p}^{\text T}\bm{x}_{r,j}$\\
		\;\;\;\;\;\;\;\;\;\;\;\;\;\;\;\;\;\;\;\;\;\;\;\;\;\;\;\;$\bm{x}_{r,j,1}=\big[\bm{x}_{r,j,1,1}^{\text T},\bm{x}_{r,j,1,2}^{\text T},...,\bm{x}_{r,j,1,P}^{\text T}\big]^{\text T}$\\
		\;\;\;\;\;\;\;\;\;\;\;\;\;\;\;\;\;\;\;\;\;\;\;\;\;\;\;\;$e_{r,j,2}=d_{r,j}-\hat{\bm m}_{r,2}^{\text T}\bm{x}_{r,j,1}$\\
		\;\;\;\;\;\;\;\;\;\;\;\;\;\;\;\;\;\;\;\;\;\;\bf{end}\\
		\;\;\;\;\;\;\;\;\;\;\;\;\;\;\;\;\;\;\;\;\;\;$\hat{\bm m}_{r+k,2}=\hat{\bm m}_{r,2}+\mu_2\sum_{j=1}^N\frac{\bm{x}_{r,j,1}e_{r,j,2}}{\lvert\lvert\bm{x}_{r,j,1}\lvert\lvert_2^2+\delta}$\\
		\;\;\;\;\;\;\;\;\;\;\;\;\;\;\;\;\;\;\;\;\;\;$\hat{\bm m}_r=\sum_{p=1}^P\hat{\bm m}_{r,2,p}\otimes\hat{\bm m}_{r,1,p}$\\
		\;\;\;\;\;\;\;\;\bf else\\
		\;\;\;\;\;\;\;\;\;\;\;\;\;\;\;$\hat{\bm m}_{r+k}=\hat{\bm m}_{r}$\\
		\;\;\;\;\;\;\;\;\bf end\\
		\bf end\\
		\hline 
	\end{tabular}
\end{table}

To make the deployment of the paper more concise, here we directly extend the NSAF-NKP-II learning rule to nonlinear subband NKP scenarios, and the implementation of the TFLN-NKP-NSAF and Volterra-NKP-NSAF algorithms are shown in Table IV.

\textbf{Remark 5.} In the nonlinear system identification, the computational complexity of the proposed TFLN-NKP-NSAF and Volterra-NKP-NSAF algorithms shares the same form as that of the linear NSAF-NKP-II algorithm, i.e., $PD+4PND+3NP(D_1+D_2)+4N+(D+1)LN$. The only difference resides in the length of the input vector $\bm{x}_r$ , which has changed from $D$ to $B$. Therefore, for the TFLN-NKP-NSAF algorithm, the length of $D$ is $(2A+1)B$. Regarding the Volterra-NKP-NSAF algorithm, the length of $D$ is $(B^2+3B)/2$.

\textbf{Remark 6.} To highlight the differences between the proposed subband NKP algorithms and the existing NKP algorithms, we have provided Table III, which summarizes the key characteristics of the related methods and our proposed methods. Clearly, the proposed algorithms based on the subband structure achieve a faster convergence rate than the NLMS-NKP, NKP-GMCC, and NKP-GHSAF algorithms under correlated input. Although the RLS-NKP and APA-NKP algorithms exhibit similar convergence performance to that of the proposed NSAF-NKP-II algorithm, these two algorithms involve matrix inversion operations, resulting in higher computational costs. In non-Gaussian scenarios, since the RNSAF-NKP-LC and RNSAF-NKP-MCC algorithms inherit the inherent characteristics of the subband structure, their performance is superior to that of the NKP-GMCC and NKP-GHSAF algorithms. In nonlinear system identification scenarios, the proposed Volterra-NKP-NSAF and TFLN-NKP-NSAF algorithms also achieve faster convergence than the TFLN-NKP-NLMS and Volterra-NKP-NLMS algorithms. Overall, the performance of the proposed subband-based NKP algorithms is superior to that of the competing algorithms in applications including system identification, echo cancellation, nonlinear systems, and impulsive noise environments.
\begin{figure}[htp]
	\centering  
	\includegraphics[scale=0.42] {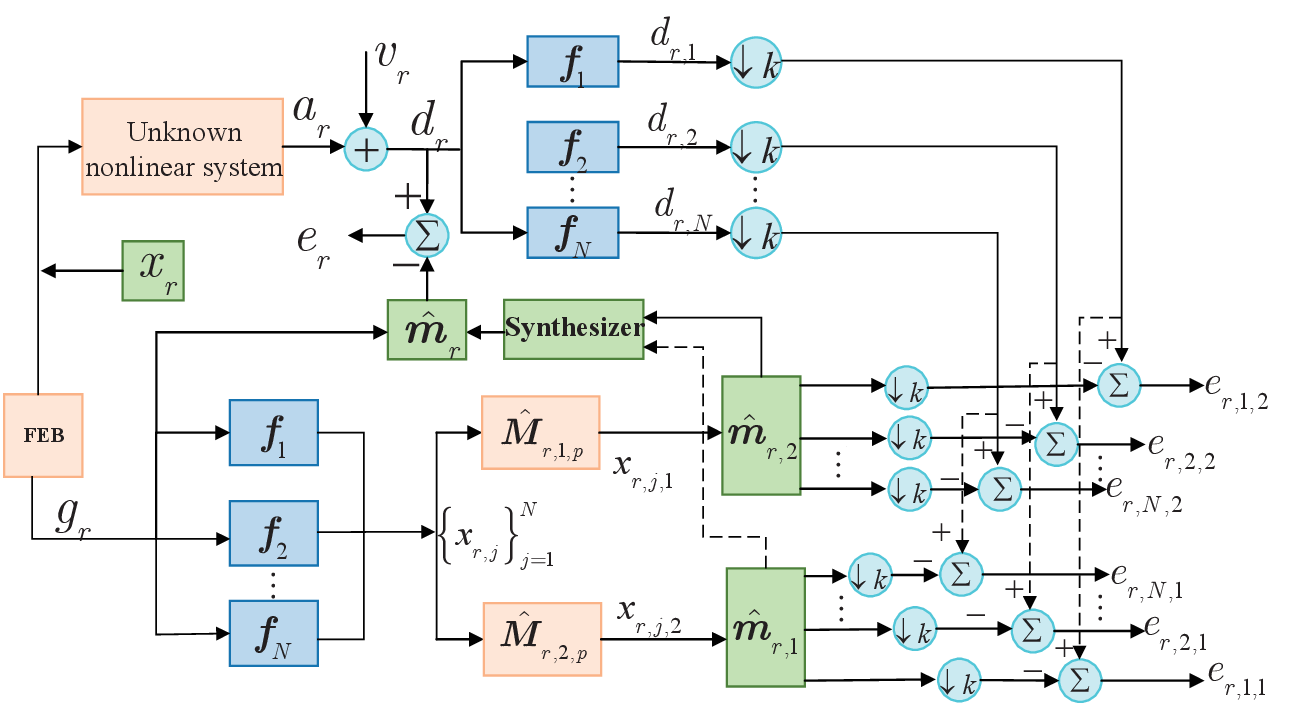}  
	\caption{Structure of nonlinear subband NKP algorithms. }
	\label{Fig16}
\end{figure}
\begin{table*}[htp]
	\renewcommand\arraystretch{0.85}
	\tabcolsep = 0.01cm
	\scriptsize
	\setlength{\abovecaptionskip}{0cm}
	\setlength{\belowcaptionskip}{0cm}
	\begin{center}
		\caption{Simulation parameters of the proposed algorithm in different application scenarios.}
		\begin{tabular}{c c c c c c c c c} 
			\hline
			\rowcolor{gray!30}	 &US-1&US-2&US-2& AEC&AEC&AEC&Nonlinear& \\	
			\rowcolor{gray!30}	 Algorithms  &Gaussian&Gaussian&$\alpha$-stable noise& Gausian&Gaussian&$\alpha$-stable noise&Case 1&ANC \\
			\rowcolor{gray!30}	  &&&&US-1&US-2&US-2&Case 2& \\			
			\hline
			 &$D_1$=25,$D_2$=20&$\mu_1$=$\mu_2$=0.02&&$D_1$=25,$D_2$=20&$\mu_1$=$\mu_2$=0.02&&&	 \\
			NSAF-NKP-I &$\mu_1$=$\mu_2$=0.03&$D_1$=25,$D_2$=20&/&$\mu_1$=$\mu_2$=0.03&$D_1$=25,$D_2$=20&/&/&/	 \\
			 &$\lambda$=0.01,$P$=2,$k$=4&$\lambda$=0.01,$P$=2,$k$=4&&$\lambda$=0.01,$P$=2,$k$=4&$\lambda$=0.01,$P$=2,$k$=4&&& \\
			\rowcolor{gray!30} &$D_1$=25,$D_2$=20&$D_1$=25,$D_2$=20&&$D_1$=25,$D_2$=20&$D_1$=25,$D_2$=20&&$D_1$=10,$D_2$=5&	 \\
			\rowcolor{gray!30}NSAF-NKP-II&$\mu_1$=$\mu_2$=0.03&$\mu_1$=$\mu_2$=0.02&/&$\mu_1$=$\mu_2$=0.03&$\mu_1$=$\mu_2$=0.02&/&$\mu_1$=$\mu_2$=0.01&/	 \\
			\rowcolor{gray!30}&$\lambda$=0.01,$P$=2,$k$=4&$\lambda$=0.01,$P$=2,$k$=4&&$\lambda$=0.01,$P$=2,$k$=4&$\lambda$=0.01,$P$=2,$k$=4&&$\lambda$=0.1,$P$=2,$k$=4&\\
			 &&$\mu_1$=$\mu_2$=0.02,$\beta$=1&$\mu_1$=$\mu_2$=0.02,$\beta$=1&&&$\mu_1$=$\mu_2$=0.08,$\beta$=20&&$\mu_1$=$\mu_2$=0.01,$k$=4	 \\
			RNSAF-NKP-LC &/&$D_1$=25,$D_2$=20&$D_1$=25,$D_2$=20&/&/&$D_1$=25,$D_2$=20&/&$\beta$=0.2,$D_1$=$D_2$=10	 \\
			  &&$\lambda$=0.01,$P$=2,$k$=4&$\lambda$=0.01,$P$=2,$k$=4&&&$\lambda$=0.01,$P$=2,$k$=4&&$\lambda$=0.01,$P$=2	 \\
			\rowcolor{gray!30} &&$\mu_1$=$\mu_2$=0.02,$\psi$=1&$\mu_1$=$\mu_2$=0.02,$\psi$=5&&&$\mu_1$=$\mu_2$=0.08,$\psi$=20&&$\mu_1$=$\mu_2$=0.01,$k$=4	 \\
			\rowcolor{gray!30}RNSAF-NKP-MCC &/&$D_1$=25,$D_2$=20&$D_1$=25,$D_2$=20&/&/&$D_1$=25,$D_2$=20&/&$\psi$=0.2,$D_1$=$D_2$=10	 \\
			\rowcolor{gray!30} &&$\lambda$=0.01,$P$=2,$k$=4&$\lambda$=0.01,$P$=2,$k$=4&&&$\lambda$=0.01,$P$=2,$k$=4&&$\lambda$=0.01,$P$=2	 \\
			 &&&&&&&$\mu_1$=$\mu_2$=0.01,$A$=2,$P$=2,$k$=4	& \\
			TFLN-NKP-NSAF &/&/&/&/&/&/&$B$=10,$D_1$=10,$D_2$=5,$\lambda$=0.1	&/ \\
			\rowcolor{gray!30} &&&&&&&$B$=10,$D_1$=13,$D_2$=5, $\lambda$=0.1&	 \\
			\rowcolor{gray!30}Volterra-NKP-NSAF &/&/&/&/&/&/&$\mu_1$=$\mu_2$=0.01,$P$=2,$k$=4&/	 \\
		    &&&&&&&	&$D_1$=$D_2$=10,$P$=2,$k$=4 \\
		     NKP-FxNSAF &/&/&/&/&/&/&/	&$\mu_1$=$\mu_2$=0.01,$\lambda$=0.01 \\
			\hline
		\end{tabular}
	\end{center}
\end{table*}
\section{Simulations}
This section validates our algorithms' performance in sparse system identification via comprehensive simulations. Based on the cosine-modulated principle, the analysis filters $\{\bm{f}_i\}_{i=1}^N$ with subband number $N=4$ and length $L=33$ are designed based on the prototype filter \cite{lee2004}. Fig. 5 describes the impulse responses of two sparse systems $\bm{m}_{0}$ of length $D=500$. In the simulation experiments, we refer to the impulse responses in Figs. 6(a) and (b) as the unknown system 1 (US-1) and unknown system 2 (US-2), respectively. The primary parameter settings for the proposed algorithms across different application scenarios are provided in Table V. Unless otherwise specified, the simulation results are averaged over 50 Monte Carlo trials.
\begin{figure}[htp]
\centering  
\includegraphics[scale=0.3] {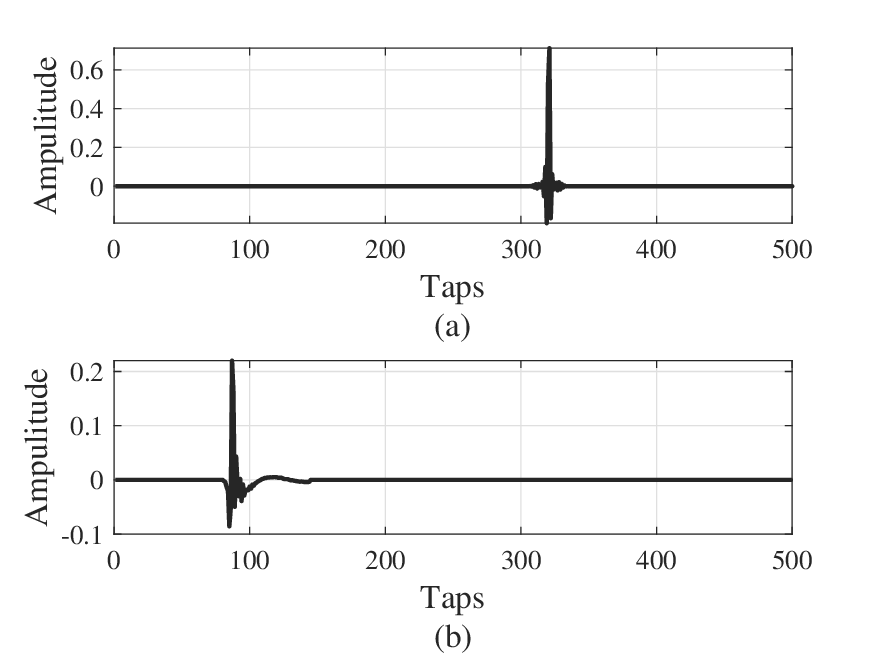}  							  
\caption{Impulse response of the unknown system $\bm{m}_0$ with $D=500$. (a) acoustic echo path \cite{zhao2014memory} and (b) network echo path is chosen from G.168 Recommendation \cite{ITU}.}
\label{Fig5}
\end{figure}
\subsection{System identification}
In the system identification scenario, the input signal $x_r$ is generated by a zero-mean Gaussian noise $w_r$ through a first-order auto-regressive (AR(1)) model with a pole at 0.9. Unless otherwise specified, the additive noise $v_r$ is a white Gaussian noise with variance $\sigma_{v}^2=0.01$. We employ the normalized mean-square deviation (NMSD), mean-square error (MSE), and EMSE to evaluate the performance of the algorithms, which is defined as
\begin{equation}
	\begin{split}
		\begin{array}{rcl}
			\begin{aligned}
				\label{057}
				\ \text{NMSD}(\text{dB})\overset{\bigtriangleup}{=}20\text{log}_{10}\text{E}\Big\{\frac{\lvert\lvert \bm{m}_0-\hat{\bm m}_r\lvert\lvert_2}{\lvert\lvert \bm{m}_0\lvert\lvert_2}\Big\},
			\end{aligned}
		\end{array}
	\end{split}
\end{equation}
\begin{equation}
	\begin{split}
		\begin{array}{rcl}
			\begin{aligned}
				\label{058}
				\ \text{MSE}(\text{dB})\overset{\bigtriangleup}{=}20\text{log}_{10}\text{E}\Big\{ e_r\Big\},
			\end{aligned}
		\end{array}
	\end{split}
\end{equation}
\begin{equation}
	\begin{split}
		\begin{array}{rcl}
			\begin{aligned}
				\label{058_1}
				\ \text{EMSE}(\text{dB})\overset{\bigtriangleup}{=}20\text{log}_{10}\text{E}\Big\{\bm{x}_r^{\text T}\widetilde{\bm m}_r\Big\}.
			\end{aligned}
		\end{array}
	\end{split}
\end{equation}

$Case $ A) Verification of Theoretical Results

The input signal is a zero-mean Gaussian noise. As shown in Fig. 7(a), we utilize the theoretical model \eqref{A20} to estimate the steady-state EMSE of the NSAF-NKP-II algorithm. Clearly, when $\mu=0.2$ and $\mu=0.5$, the theoretical steady-state EMSE model can accurately predict the simulation results of the algorithm. However, when $\mu=0.7$, the theoretical steady-state EMSE value fails to closely match the simulation value. Therefore, a more precise theoretical steady-state model deserves further investigation by researchers. As shown in Fig. 7(b), the step-size $\mu=1$ exceeded the range specified in \eqref{A22}, which caused the NSAF-NKP-II algorithm to diverge.
\begin{figure}[htp]
	\centering  
	\includegraphics[scale=0.29] {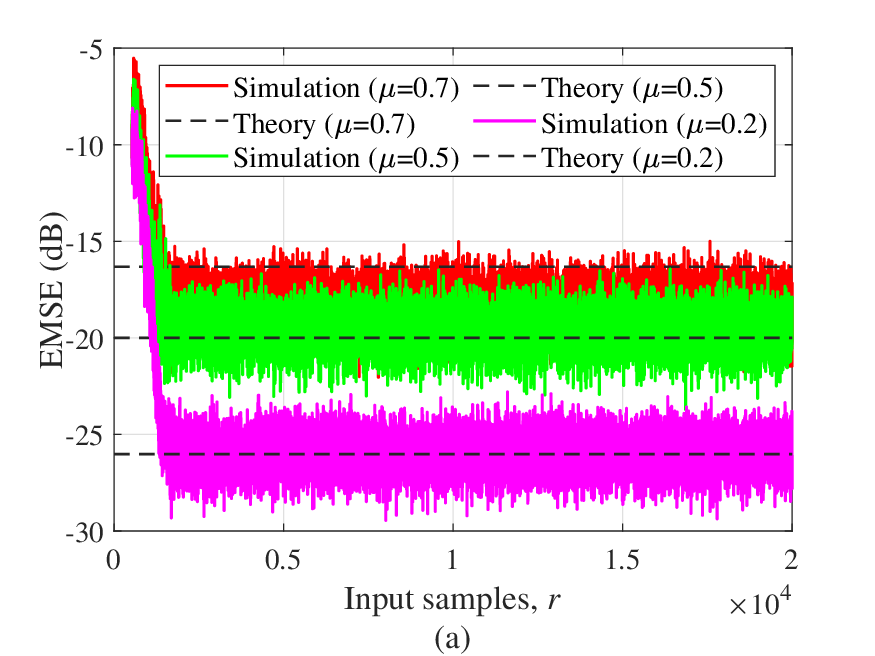}  
	\hspace{0.001ex}	
	\includegraphics[scale=0.29] {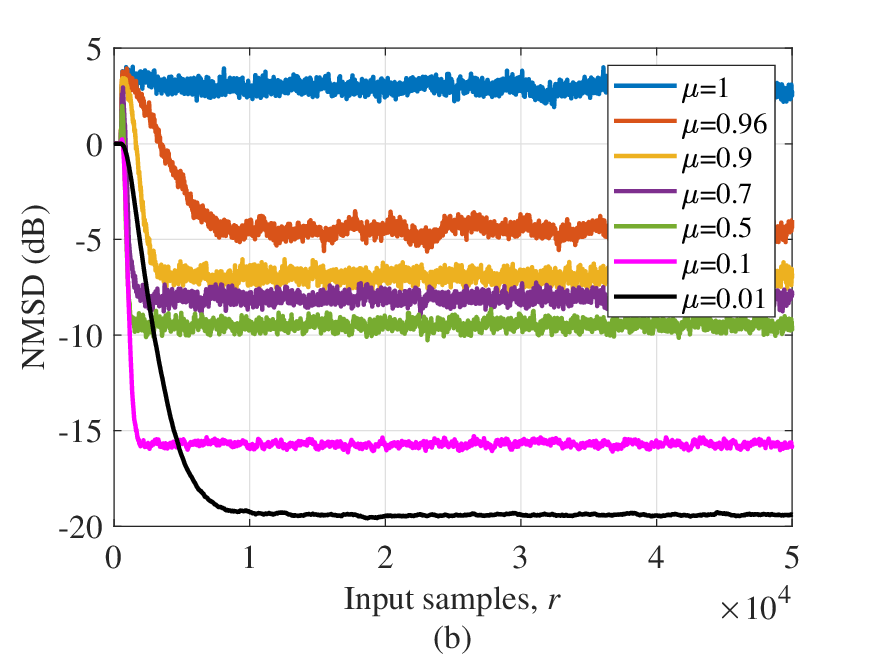}  
	\hspace{0.001ex}									 									  
	\caption{(a) Theoretical and simulated MSDs versus; (b) NMSDs curves of the NSAF-NKP-II algorithm with different step-sizes.}
	\label{Fig6}
\end{figure}

$Case $ B) Influence of parameters

Fig. 8 analyzes the influence of different parameters (e.g., $N$, $L$, $k$, $P$, $\lambda$, $\psi$, and $\beta$) on the performance of the proposed subband NKP algorithms during the identification of US-2. In Fig. 8(a), we analyze the impact on the performance of the NSAF-NKP-II algorithm when the sub-dimensions $D_1$ and $D_2$ do not match the actual system dimensions $D$ (i.e., $D_1\times D_2\neq D$). Clearly, the performance of the NSAF-NKP-II algorithm with $D_1\times D_2=400$ and $D_1\times D_2=200$ is basically the same as that of the NSAF-NKP-II algorithm with $D_1\times D_2=500$. However, the case of $D_1\times D_2=100$ is unable to fully describe the characteristics of the system to be identified, which leads to a deterioration in the steady-state MSE performance of the NSAF-NKP-II algorithm. Notably, we usually set $D_1\times D_2=D$ to ensure that the adaptive filter has sufficient degrees of freedom to accurately model the system to be identified in the system identification scenarios.
\begin{figure}[htp]
	\centering  
	\includegraphics[scale=0.29] {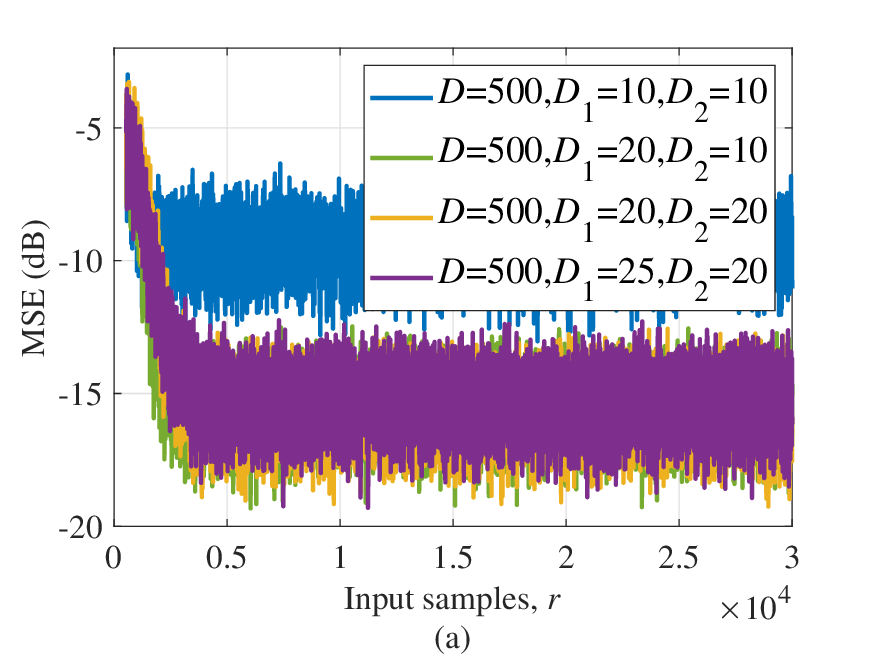}
	\hspace{0.001ex}
	\includegraphics[scale=0.29] {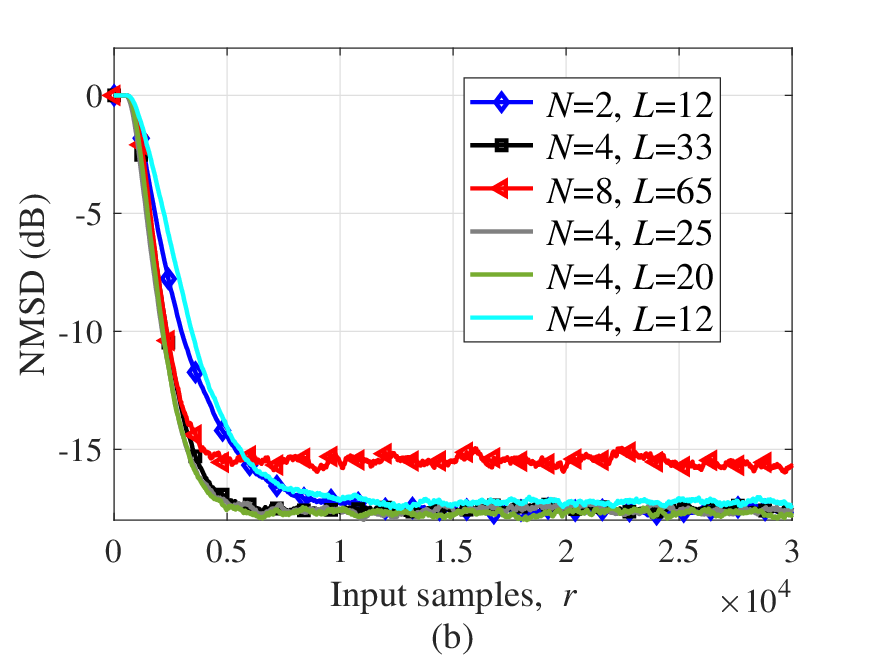}  
	\hspace{0.001ex}									 
	\includegraphics[scale=0.29] {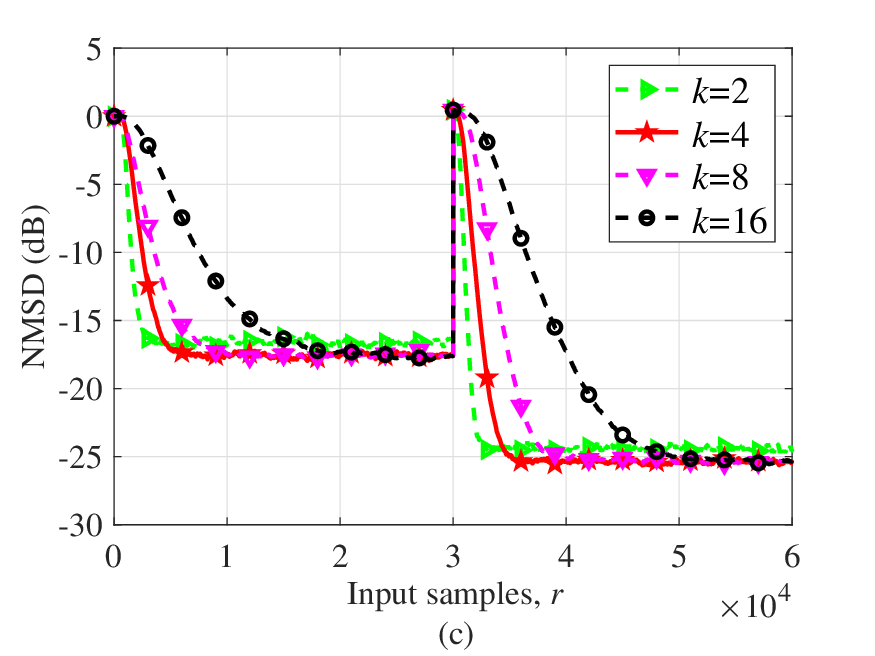}	
	\hspace{0.001ex}
	\includegraphics[scale=0.29] {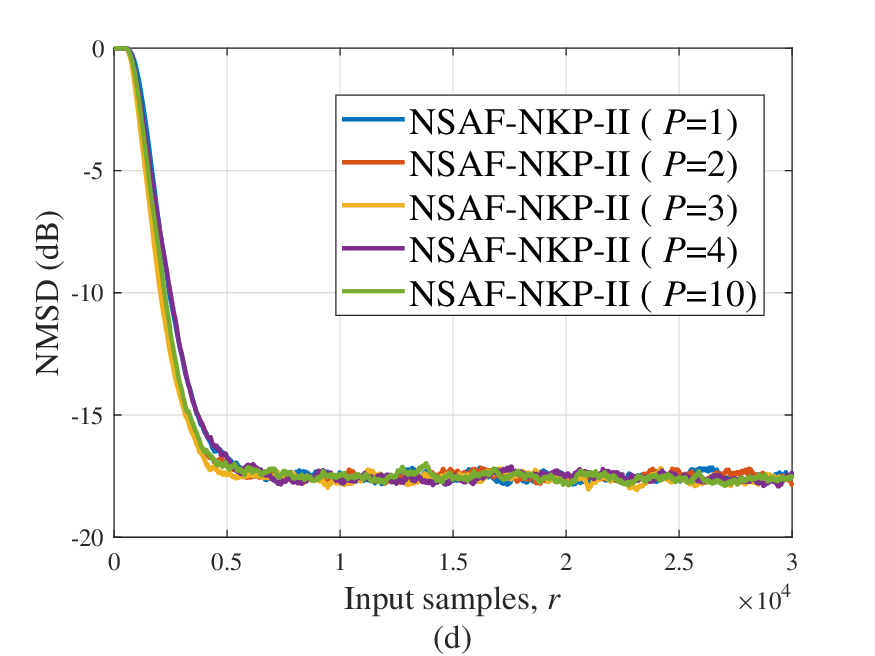}	
	\hspace{0.001ex}
	\includegraphics[scale=0.29]{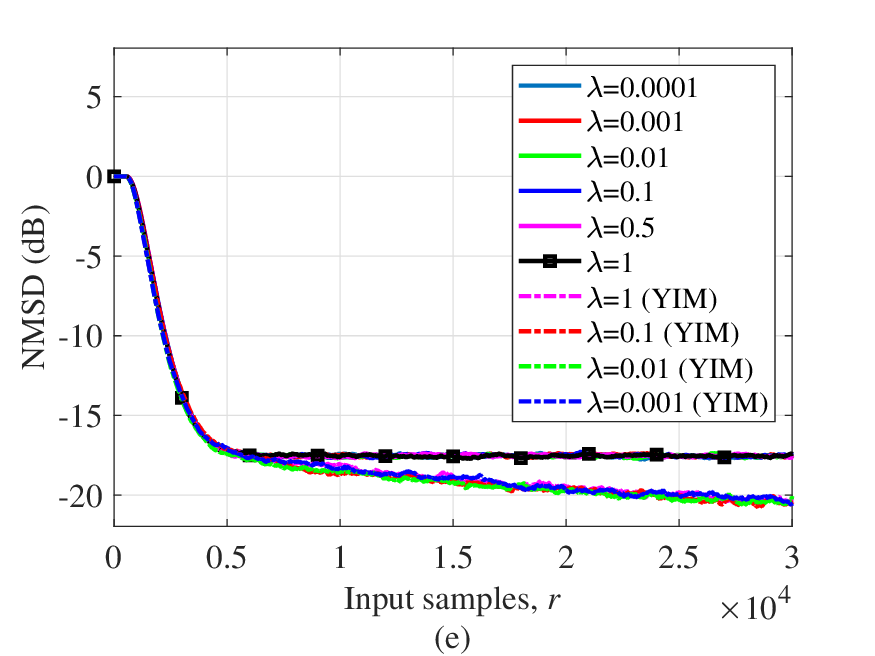}							
	\hspace{0.001ex}
	\includegraphics[scale=0.29]{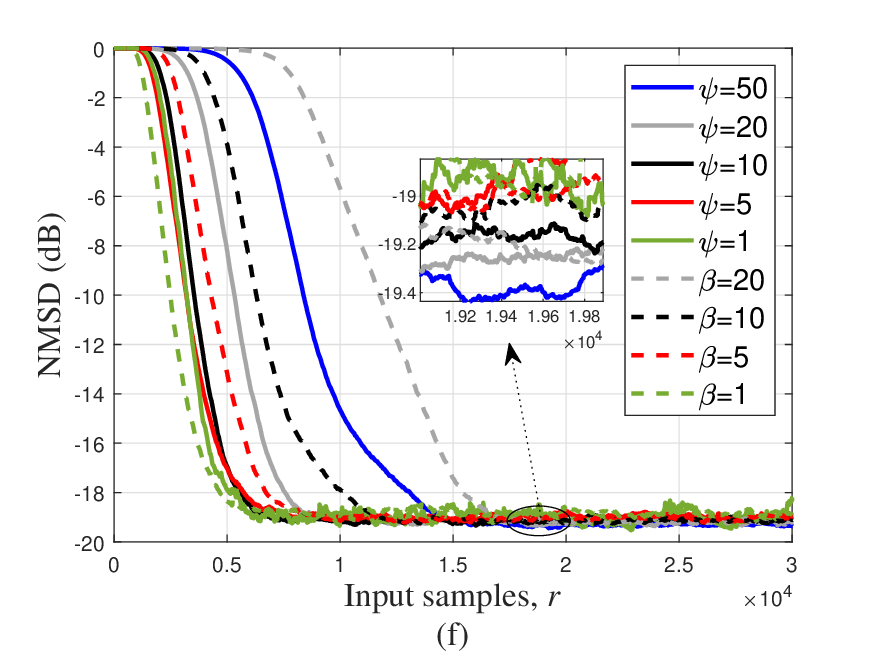}	  
	\caption{NMSD curves of the NSAF-NKP-II algorithm for different parameters. }
	\label{Fig6}
\end{figure}

Fig. 8(b) investigates the impact of different parameters $N$ and $L$ on the performance of the NSAF-NKP-II algorithm. As the number of subbands $N$ increases, the filter's ability to decorrelate correlated inputs becomes stronger, thereby accelerating the convergence rate of the NSAF-NKP-II algorithm. However, if the number of subbands exceeds $N=4$ in this experiment, the convergence rate of the algorithm will remain basically unchanged, but the steady-state performance will deteriorate. Therefore, we set $N=4$ in the following examples. In addition, when $N=4$, a slightly smaller $L$ (such as $L=25$ and $L=20$) does not adversely affect the performance of the NSAF-NKP-II algorithm. However, the value of $L$ cannot be too small, as doing so will slow down the convergence rate of the algorithm.

Fig. 8(c) depicts the NMSD performance of the NSAF-NKP-II algorithm for different $k$. Clearly, the smaller the update interval $k$, the more significantly the convergence rate of the NSAF-NKP-II algorithm is improved, but the computational complexity is also correspondingly increased. Notably, when the number of $k$ becomes smaller than $N$, the improvement in the convergence rate of the NSAF-NKP-II algorithm becomes insignificant, and the steady-state NMSD becomes larger. Therefore, we set $k=4$ in the following examples. Furthermore, we investigate the algorithm's tracking efficacy under dynamic system conditions by introducing an abrupt transition (from US-2 to US-1) during the middle of the input sample. The numerical results substantiate that the NSAF-NKP-II algorithm maintains robust tracking capability, demonstrating swift adaptation to sudden parameter variations in time-varying environments.

Figs. 8(d) and (e) analyze the parameter $P$ and the initial value $\lambda$ on the performance of NSAF-NKP-II algorithm. The proposed NSAF-NKP-II algorithm is not affected by the parameters $P$ and $\lambda$ in the current experiment scenario. In addition, the NSAF-NKP-II algorithm based on the YIM method achieves a lower steady-state error than the original method. However, to be consistent with other NKP adaptive filtering algorithms, we still choose to use the original method. 

Fig. 8(f) illustrates the NMSD performance of the RNSAF-NKP-LC and RNSAF-NKP-MCC algorithms with varying $\beta$ and $\psi$ parameter values, respectively. As evident from the results, the larger the values of $\beta$ and $\psi$, the lower the steady-state errors of the RNSAF-NKP-LC and RNSAF-NKP-MCC algorithms become. However, their values cannot be too large, because we also need to consider the convergence rate of gradient-based adaptive methods.

In Fig. 9, we check the performance of the NSAF-NKP-II, RNSAF-NKP-MCC, and RNSAF-NKP-LC algorithms in Gaussian noise under US-2. Clearly, compared with the NSAF-NKP-II algorithm, the performance of the robust subband NKP algorithms in terms of convergence rate and steady-state NMSD is not compromised under the Gaussian noise scenario, even with the introduction of the corresponding robustness criteria.

\begin{figure}[htp]
	\centering  
	\includegraphics[scale=0.3] {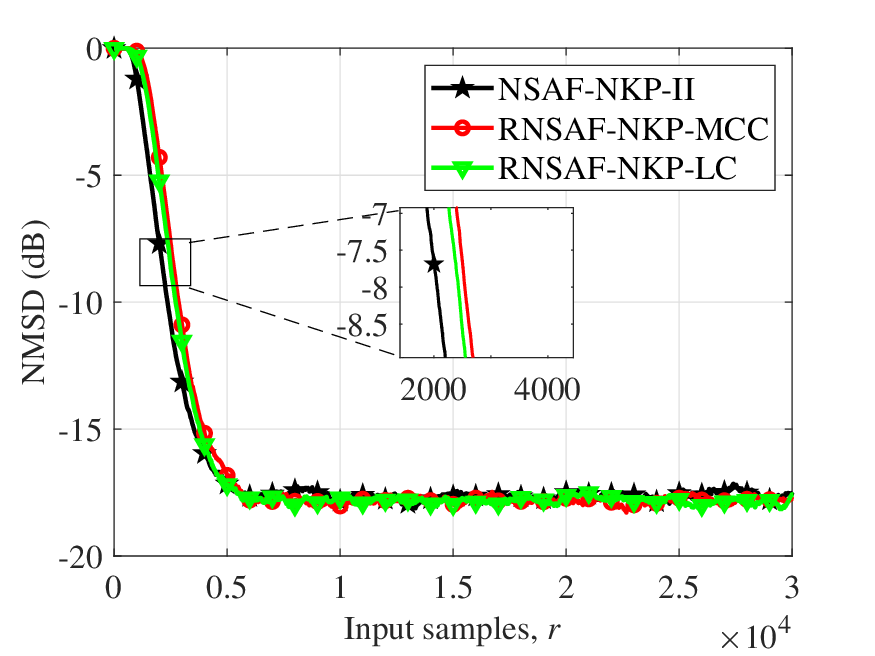}  
	\caption{NMSD curves of the NSAF-NKP-II, RNSAF-NKP-MCC, and RNSAF-NKP-LC algorithms in the Gaissian noise. }
	\label{Fig11}
\end{figure}
\begin{figure}[htp]
	\centering  
	\includegraphics[scale=0.29] {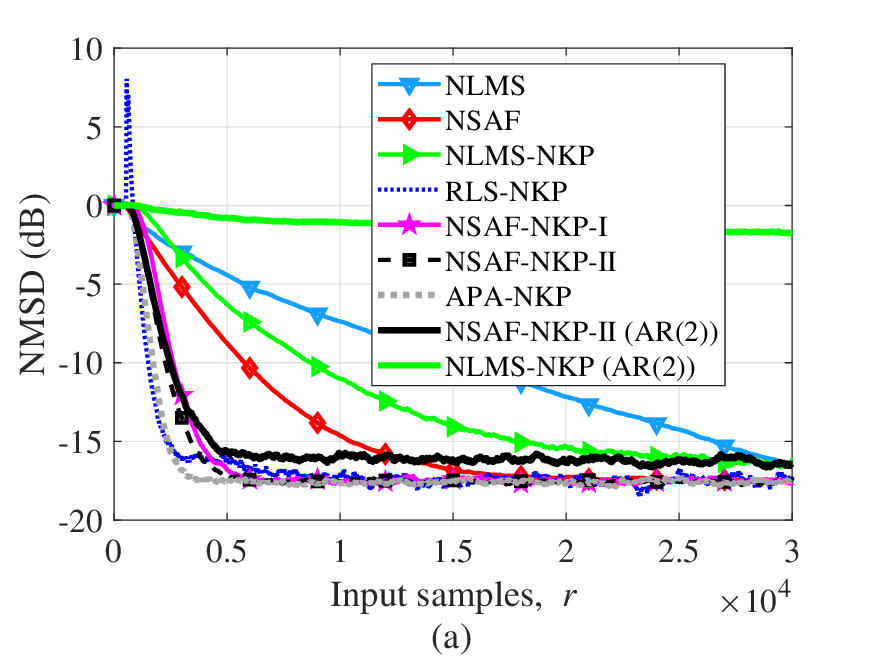}  
	\hspace{0.001ex}									 
	\includegraphics[scale=0.29] {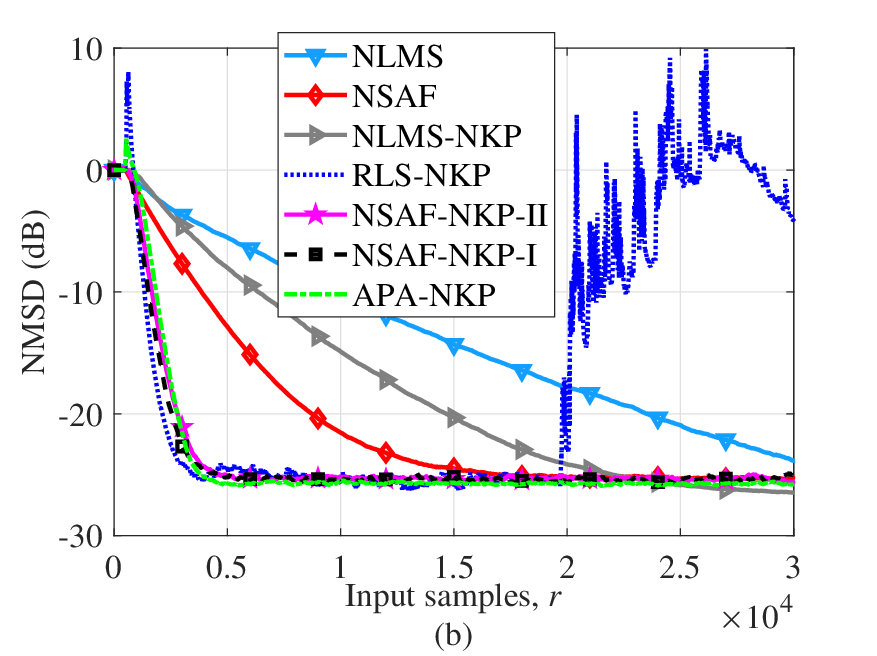}								  
	\caption{(a) NMSD curves of the algorithms in the identification of US-2, [$P=2$, $\mu_1=\mu_2=0.02$] for the NLMS-NKP algorithm; [$\lambda_1=0.998$, $\lambda_2=0.998$, $P=2$] for the RLS-NKP algorithm, [$P=2$, $C=5$, $\mu_1=\mu_2=0.008$] for the APA-NKP; [$\mu=0.02$] for the NLMS and NSAF algorithms; (b) NMSD curves of the algorithms in the identification of US-1, [$\mu=0.03$] for the NLMS-NKP algorithm; [$\mu_1=\mu_2=0.009$] for the APA-NKP; [$\mu=0.03$] for the NLMS and NSAF algorithms, the other parameters of the algorithms are the same as those in Fig. 10(a). }
	\label{Fig7}
\end{figure}

$Case $ C) Gaussian noise scenario

Fig. 10(a) compares the NMSD curves of the NLMS \cite{slock1993}, NSAF \cite{lee2004}, NLMS-NKP \cite{bhattacharjee2020}, RLS-NKP \cite{elisei2019}, APA-NKP \cite{li2025nearest}, NSAF-NKP-I, and NSAF-NKP-II algorithms for identifying the US-2 system. In this figure, the second auto-regressive (AR(2)) input generated by the equation $x_r=1.5x_{r-1}-0.6x_{r-2}+w_r$. Notably, AP-based NKP algorithms have not been previously studied in the context of system identification. To ensure a fair comparison, we extended the NKP-MFxAP algorithm from \cite{li2025nearest} to the system identification scenario, resulting in the NKP-APA algorithm. Comparative analysis reveals that the NSAF algorithm demonstrates improved convergence performance relative to the NLMS algorithm, attributable to its SAF structure. Notably, the NLMS-NKP algorithm exhibits superior initial convergence characteristics compared to the NLMS algorithm. The proposed NSAF-NKP-I algorithm further improves the convergence rate over the NLMS-NKP method, albeit at the expense of increased multiplicative computational load. Fortunately, the proposed NSAF-NKP-II algorithm achieves comparable convergence performance to that of NSAF-NKP-I while substantially reducing computational complexity. Although the RLS-NKP and APA-NKP algorithms demonstrate marginally faster convergence than the proposed algorithms, this advantage is offset by their elevated computational demands. When the correlation of the input signal is further enhanced, i.e., when the input is an AR (2) signal, the performance of the NLMS-NKP algorithm deteriorates sharply. However, the performance of the proposed NSAF-NKP-II algorithm remains largely consistent with that observed when the input is an AR(1) signal. Additionally, the fixed-order RLS-NKP algorithm is prone to numerical instability in finite precision implementation, as evidenced by the divergent behavior observed in Fig. 10(b). Fig. 10(b) compares the performance of the algorithms shown in Fig. 10(a) in the US-1 scenario. Clearly, we can draw some conclusions analogous to those presented in Fig. 10(a).
\begin{figure}[htp]
	\centering  
	\includegraphics[scale=0.3] {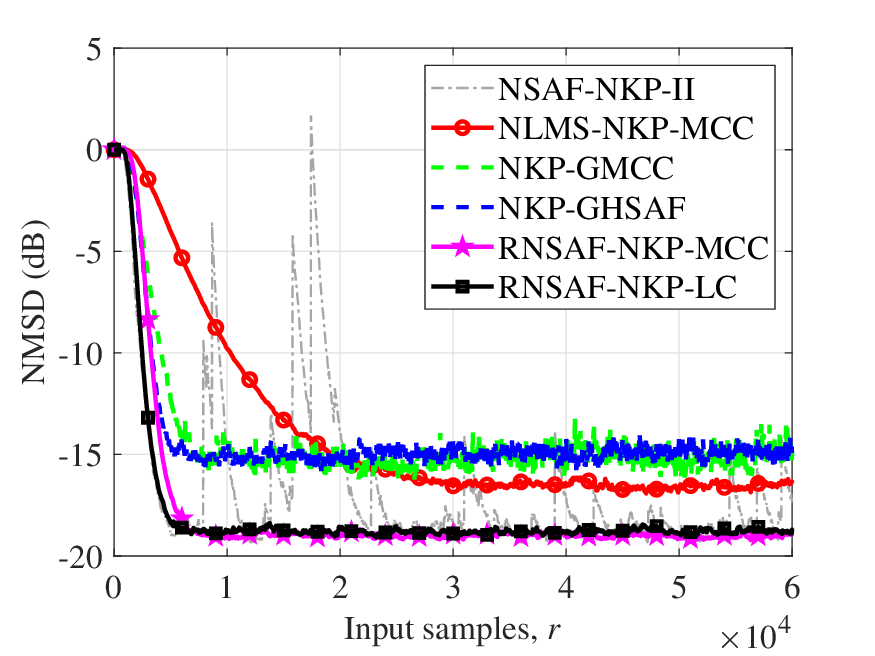}  
	\caption{NMSD curves of different algorithms for the US-2 identification in the $\alpha$-stable noise. All algorithms have $\mu_1=0.02$ and $\mu_2=0.02$. [$\alpha=2$, $\lambda=0.99$, $\mu=0.0053$] for the NKP-GHSAF algorithm; [$\alpha=2$, $\lambda=0.0002$, $\mu=0.003$] for the NKP-GMCC algorithm; [ $P=2$] for the NKP decomposition; [$\varpi=1.5$] for the $\alpha$-stable noise. }
	\label{Fig10}
\end{figure}

$Case$ D) $\alpha$-Stable Noise Scenario

The imuplsive noise $v_r$ is modeled by the $\alpha$-stable process, and its characteristic function provided as\cite{talebi2018}
\begin{equation}
	\begin{split}
		\begin{array}{rcl}
			\begin{aligned}
				\label{057_1}
				\ v_r = \text{exp}{(-\gamma\lvert r\lvert^{\varpi})},
			\end{aligned}
		\end{array}
	\end{split}
\end{equation}
where the characteristic index $\varpi\in(0,2]$ controls the impulsiveness of the noise, $\gamma>0$ governs the dispersion of noise ($\gamma=1/60$ in this paper), and the noise $v_r$ represents Gaussian noise when $\varpi=2$.
\begin{figure}[htp]
	\centering  
	\includegraphics[scale=1] {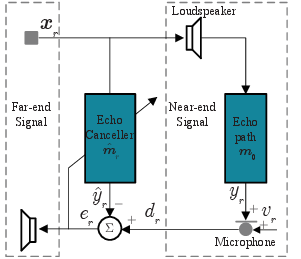}  
	\caption{Structure of the adaptive echo canceler.}
	\label{Fig14}
\end{figure}

Fig. 11 compares the NMSD curves of the NSAF-NKP-II, MCC-based NLMS-NKP (NLMS-NKP-MCC), NKP-GMCC \cite{bhattacharjee2020nearest}, NKP-GHSAF \cite{bhattacharjee2020nearest}, RNSAF-NKP-MCC, and RNSAF-NKP-LC algorithms for the identification of US-2. Notably, no prior studies have investigated the NLMS-NKP-MCC algorithm. To ensure equitable comparative analysis, we have extended the MCC technique to the existing NLMS-NKP framework. Certainly, the non-robust NSAF-NKP-II algorithm demonstrates unstable convergence behavior under impulsive noise. The NLMS-NKP-MCC algorithm achieves stable convergence in impulsive noise environments, which is attributed to the fact that the MCC criterion reduces the step-size to a very small value during impulsive noise occurrences. By utilizing the GMCC and GHSF criteria as robust cost functions, the NKP-GMCC and NKP-GHSAF algorithms can converge stably in impulsive noise environments; moreover, their convergence rates are faster than that of the NLMS-NKP-MCC algorithm, but their steady-state misadjustment values are relatively large. Interestingly, the proposed RNSAF-NKP-MCC and RNSAF-NKP-LC algorithms achieve the fastest convergence rate while having the lowest steady-state error.
\subsection{Echo cancellation}
In the echo cancellation scenarios, we will add another metric to evaluate the performance of the algorithms in the echo cancellation scenarios, i.e., Echo Return Loss Enhancement (ERLE), which is defined as \cite{10806852}
\begin{equation}
	\begin{split}
		\begin{array}{rcl}
			\begin{aligned}
				\label{057_2}
				\ \text{ERLE}(\text{dB})\overset{\bigtriangleup}{=}10\text{log}_{10}\Big(\frac{\text{E}\{d_r^2\}}{\text{E}\{e_r^2\}}\Big),
			\end{aligned}
		\end{array}
	\end{split}
\end{equation}
where the expectation is estimated by time averaging.

$Case $ A) Gaussian noise scenario
\begin{figure}[htp]
	\centering  
	\includegraphics[scale=0.29] {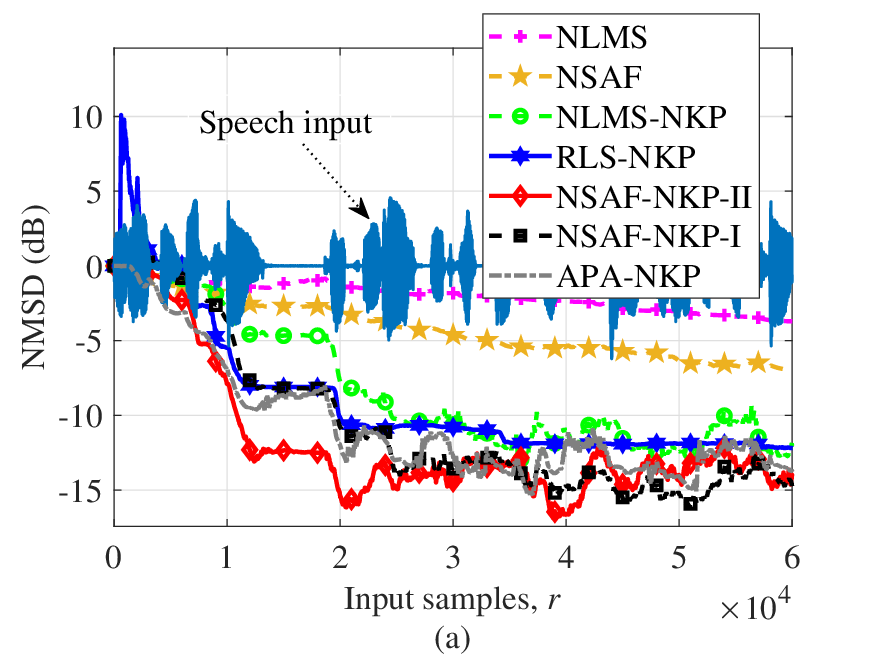}  
	\hspace{0.001ex}									 
	\includegraphics[scale=0.29] {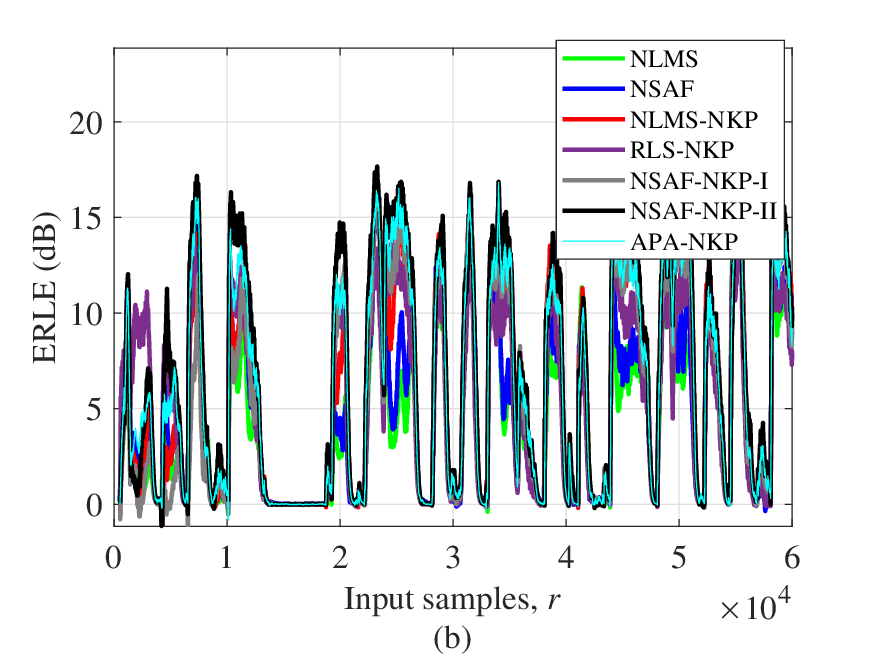}		
	\caption{NMSD and ERLE curves of the NLMS, NSAF, NLMS-NKP, RLS-NKP, APA-NKP, NSAF-NKP-I, and NSAF-NKP-II algorithms in the identification of US-2. The parameters of the algorithms are the same as those in Fig. 10. }
	\label{Fig12}
\end{figure}

For the echo cancellation system described in Fig. 12, $\bm{x}_r$ represents the speech input signal. The echo path $\bm{m}_0$ is depicted by the US-1 and US-2 systems. By putting the speech input signal $\bm{x}_r$ through the echo path, we obtain the echo signal $y_r$. Using the same input signal, the echo canceller identifies the echo path through the adaptive filter $\hat{\bm m}_r$, and its output is $\hat{y}_r$, which corresponds to the estimated value of the echo signal $y_r$. Subsequently, by subtracting $\hat{y}_r$ from the desired signal $d_r$, a higher quality speech signal $e_r$, free from echo interference, is obtained.
\begin{figure}[htp]
	\centering  
	\includegraphics[scale=0.29] {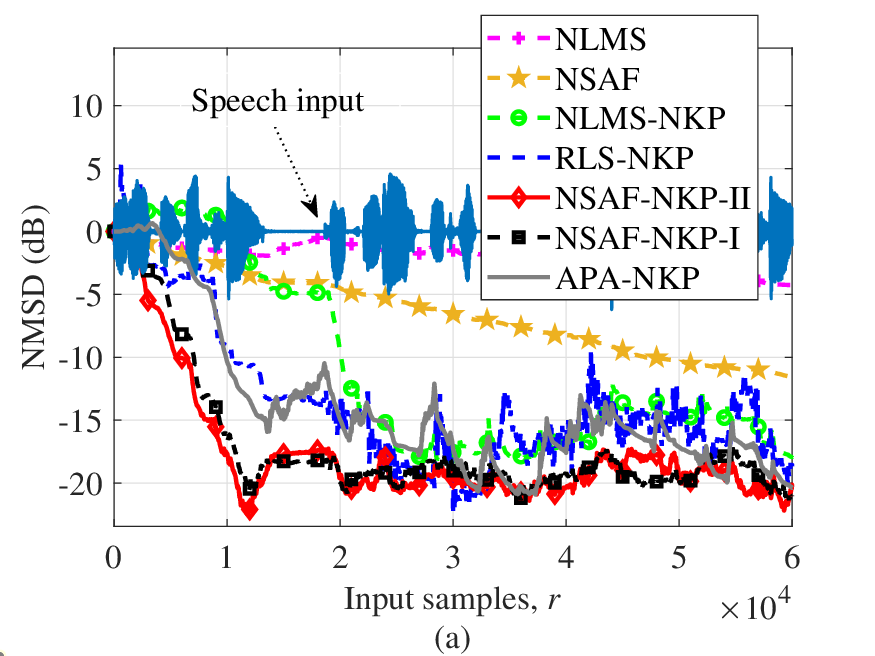} 
	\hspace{0.001ex}									 
	\includegraphics[scale=0.29] {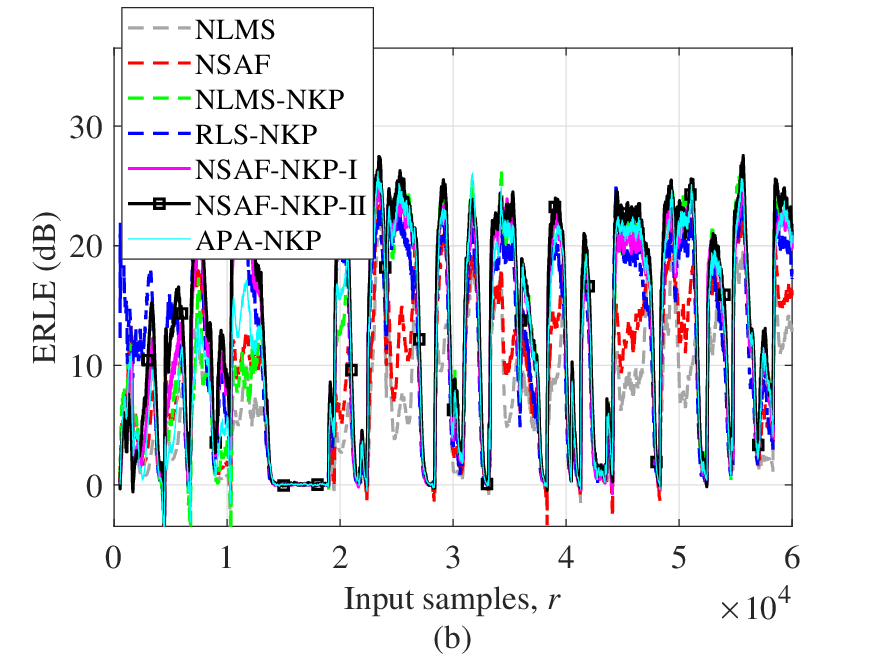}		 
	\caption{NMSD and ERLE curves of different algorithms in the identification of US-1. The parameters of the algorithms are the same as those in Fig. 10. }
	\label{Fig13}
\end{figure}

Fig. 13 compares the performance of algorithms in the identification of US-2. In Fig. 13(a), the NSAF algorithm achieves faster convergence than the NLMS algorithm. Through the application of the NKP decomposition technique, the NLMS-NKP algorithm exhibits enhanced convergence characteristics and reduced steady-state misadjustment relative to the NSAF algorithm. By updating weights using the autocorrelation matrix of the input signal, the RLS-NKP algorithm achieves superior steady-state precision compared to the NLMS-NKP algorithm. The APA-NKP algorithm, which utilizes historical input data, achieves slightly better steady-state performance compared to the RLS-NKP algorithm. The proposed NSAF-NKP-I algorithm achieves performance similar to that of the NLMS-NKP algorithm. However, the advantages of the RLS-NKP, APA-NKP, and NSAF-NKP-I algorithms are attained at the expense of increased computational complexity. Fortunately, the proposed NSAF-NKP-II algorithm addresses this limitation (i.e., it effectively reduces computational complexity), while achieving superior convergence performance. In Fig. 13(b), we employ the ERLE evaluation metric to further validate the effectiveness of the proposed algorithms, and similar conclusions can be drawn. Fig. 14 compares the performance of NLMS, NSAF, NLMS-NKP, RLS-NKP, APA-NKP, NSAF-NKP-I, and NSAF-NKP-II algorithms in the identification of US-1. We can also draw conclusions analogous to those presented in Fig. 13.
\begin{figure}[htp]
	\centering  
	\includegraphics[scale=0.29] {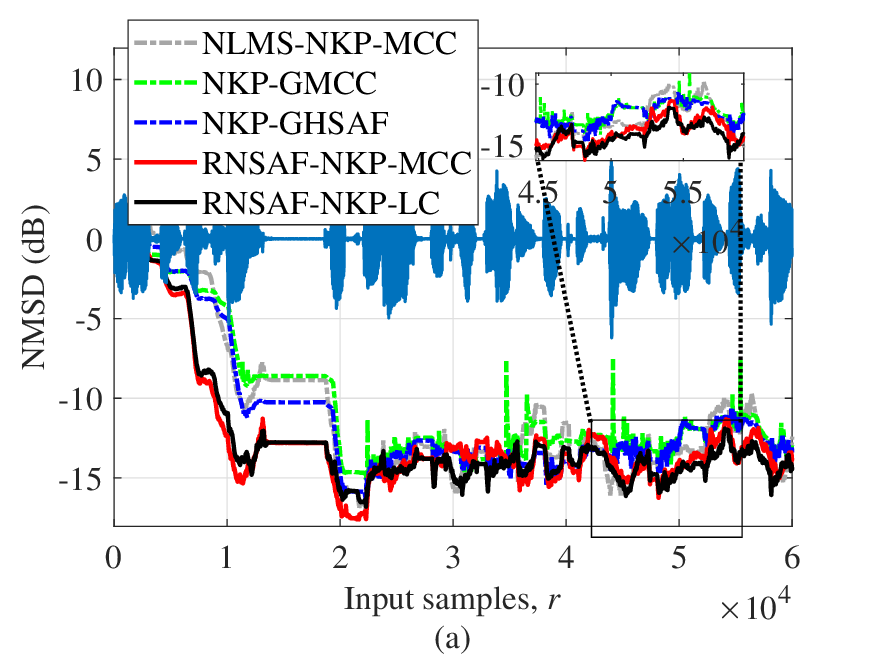} 
	\hspace{0.001ex}									 
	\includegraphics[scale=0.29] {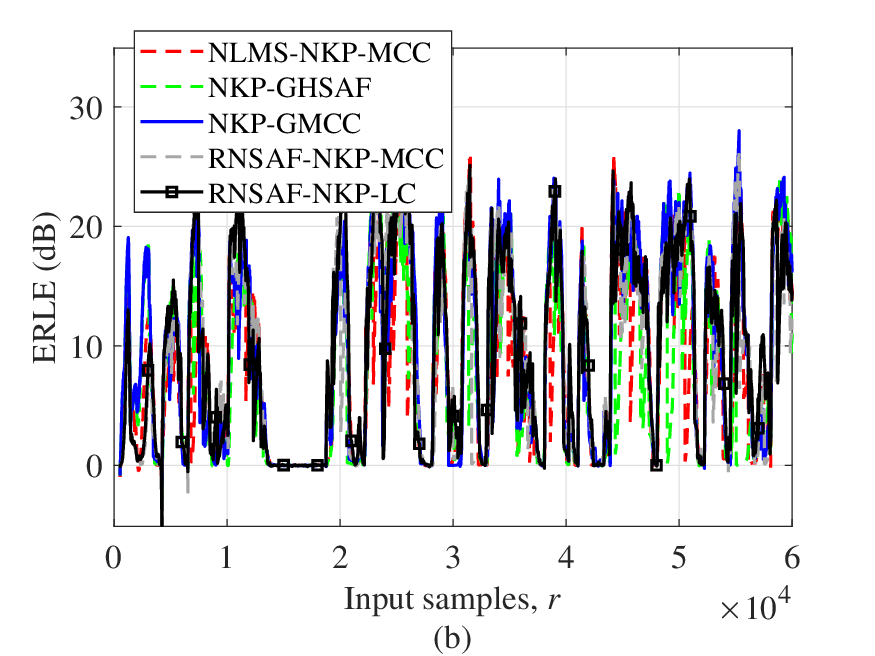}	 
	\caption{NMSD and ERLE curves of for the US-2 identification in the $\alpha$-stable noise. [$\mu_1=\mu_2=0.08$] for the NLMS-NKP-MCC algorithm; [$\alpha=2$, $\lambda=0.99$, $\mu=0.01$] for the NKP-GHSAF algorithm; [$\alpha=2$, $\lambda=0.0006$, $\mu=0.007$] for the NKP-GMCC algorithm. The parameters of the algorithms are the same as those in Fig. 11. }
	\label{Fig15}
\end{figure}

$Case$ B) $\alpha$-Stable Noise Scenario

Fig. 15 compares the proposed robust algorithms with competing algorithms in the identification of US-2. In comparison, the proposed RNSAF-NKP-MCC and RNSAF-NKP-LC algorithms show the fastest convergence and lowest steady-state misadjustment than their competitors.
\subsection{Complex nonlinear scenarios}
In the nonlinear scenarios, the input signal $x_r$ is generated by a uniform distribution random sequence within the range of $[-0.5,0.5]$, and the noise $v_r$ is a white Gaussian noise with variance $\sigma_{v}^2=0.001$. 

$Case$ A) Asymmetric loudspeaker distortion \cite{comminiello2013functional}

The nonlinear model is established as
\begin{equation}
	\begin{split}
		\begin{array}{rcl}
			\begin{aligned}
				\label{059}
				\ a_r=\phi\Big[\frac{1}{1+e^{-\vartheta a_{r,1}}}-\frac{1}{2}\Big],
			\end{aligned}
		\end{array}
	\end{split}
\end{equation}
where $a_{r,1}=1.5x_r-0.3x^2_r$, $\phi=2$, and
\begin{equation}
	\label{060}
	\left\{ \begin{aligned}
		&\vartheta=4 \text{ if } a_{r,1}>0\\ 
		&\vartheta=\frac{1}{2} \text{ if } a_{r,1}\leq 0 \nonumber.
	\end{aligned} \right.
\end{equation}

Fig. 16(a) compares the MSE curves of the NSAF-NKP-II, TFLN-NSAF \cite{zhang2023design}, TFLN-NKP-NLMS \cite{bhattacharjee2021}, TFLN-NKP-NSAF, Volterra-NKP-NSAF, Volterra-NKP-NLMS \cite{bhattacharjee2021}, and Volterra-NSAF \cite{10806852} algorithms under conditions of asymmetric loudspeaker distortion. Unfortunately, the proposed linear NSAF-NKP-II algorithm performs poorly in steady-state precision when encountering this asymmetric loudspeaker distortion. In comparison, the nonlinear algorithms based on the TFLN and Volterra structures can achieve better MSE performance in terms of convergence and steady-state. For the algorithms with a TFLN structure, the proposed TFLN-NKP-NSAF algorithm obtains a faster convergence and better steady-state precision than the TFLN-NSAF and TFLN-NKP-NSAF algorithm. Furthermore, for the algorithms with a Volterra structure, analogous conclusions can be drawn to those for the TFLN-based counterparts.
\begin{figure}[htp]
	\centering  
	\includegraphics[scale=0.29] {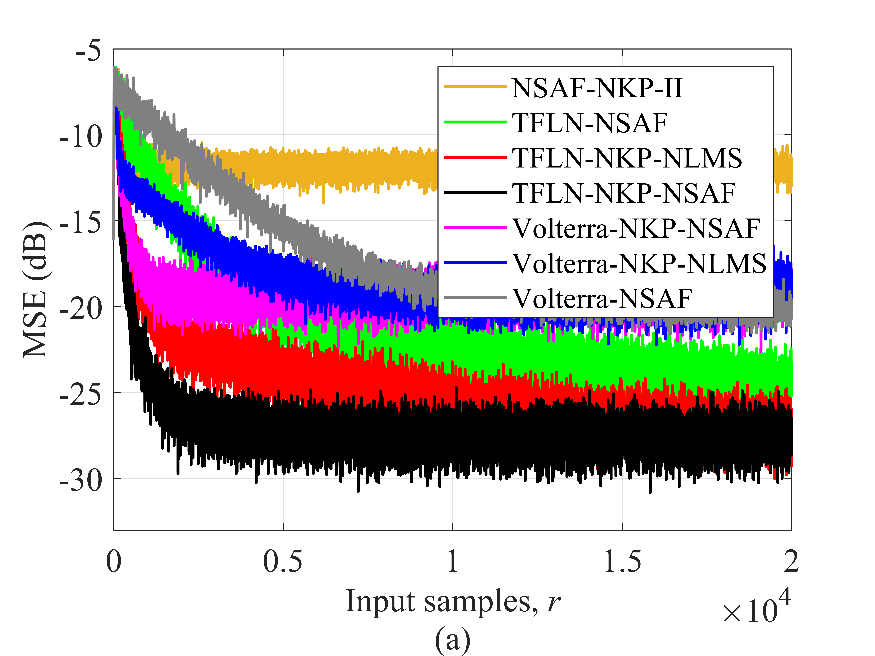}
	\hspace{0.001ex}
	\includegraphics[scale=0.29] {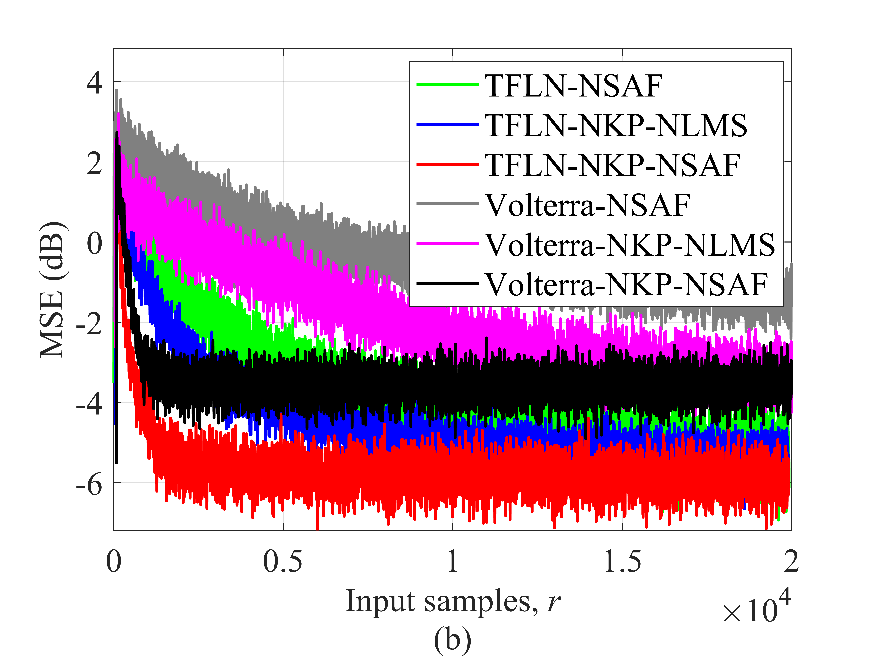}   
	\caption{Comparison of proposed algorithms with competing algorithms. The MSE curves were obtained by averaging results over 200 independent trials. [$A=2$, $B=10$, $D=50$, $D_1=10$, $D_2=5$, $\mu_1=\mu_2=0.01$, $P=2$] for the TFLN-based algorithms; [$B=10$, $D=65$, $D_1=13$, $D_2=5$, $\mu_1=\mu_2=0.01$, $P=2$] for the Volterra-based algorithms.} 
	\label{Fig17}
\end{figure}

$Case$ B) Symmetrical soft-clipping nonlinearity \cite{comminiello2014nonlinear}

The nonlinear model is given by
\begin{equation}
	\label{061}
	a_r=\left\{ \begin{aligned}
		&\frac{2}{3\tau}x_r,\;\;\;\;\;\;\;\;\;\;\;\;\;\;\;\;\;\;\;\;\;\;\;\;\;\;\;\;\;\;\; 0\leq \lvert x_r\lvert\leq\tau\\ 
		&\text{sign}[x_r]\frac{3-[2-x_r/\tau]^2}{3}, \; \tau\leq \lvert x_r\lvert\leq 2\tau\\
		&\text{sign}[x_r],\;\;\;\;\;\;\;\;\;\;\;\;\;\;\;\;\;\;\;\;\;\;\;\;\;\;\; 2\tau\leq \lvert x_r\lvert\leq1
	\end{aligned} \right.
\end{equation}
where $\tau\in(0,0.5]$ ($\tau=0.3$ in our simulation).

Fig. 16(b) compares the proposed algorithms with competing algorithms under the symmetrical soft-clipping nonlinearity environment. As depicted, the proposed TFLN-NKP-NSAF and Volterra-NKP-NLMS algorithms obtain better performance than their competing algorithms.

$Case$ C) Multiplication complexity of nonlinear algorithm

\begin{figure}[htp]
	\centering
	\includegraphics[scale=0.29] {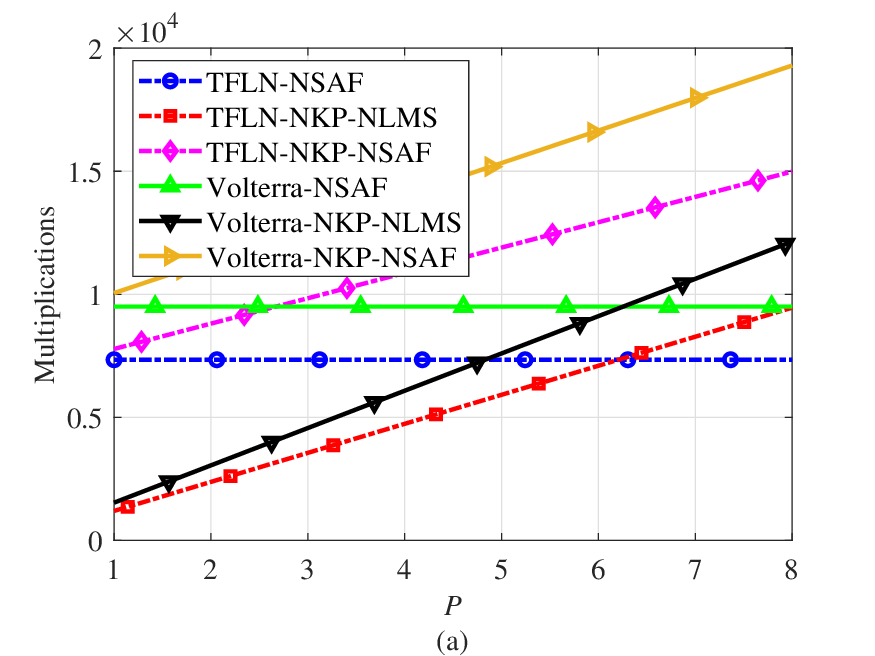}
	\hspace{0.001ex}
	\includegraphics[scale=0.29] {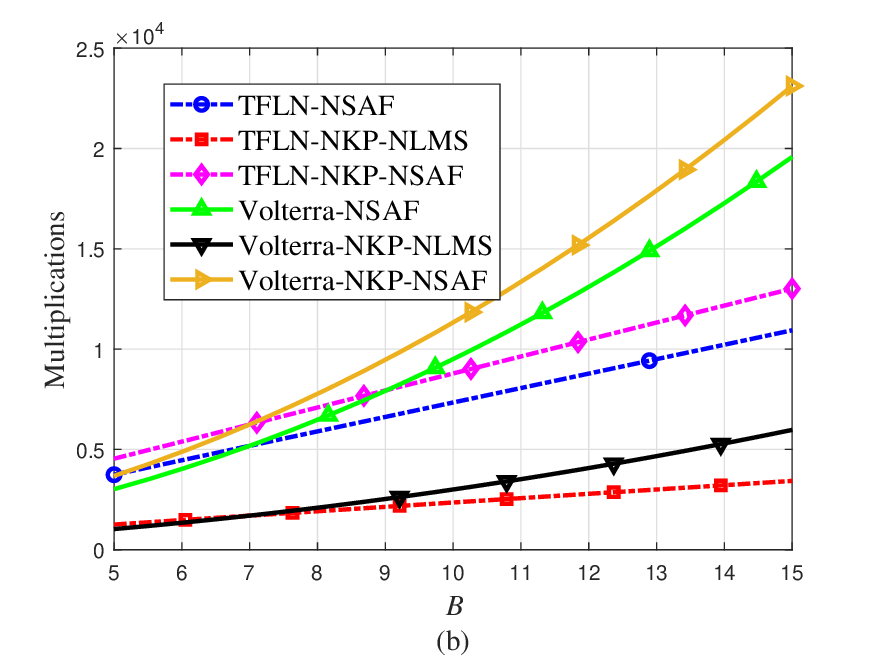} 
	\vspace{-1em} \caption{(a) The computational complexity of multiplications versus the parameter $P$; (b) The computational complexity of multiplications versus the parameter $B$. $N=4$ and $k=4$ for Fig. 17(a); $D_1$=$D_2$=$\sqrt{D}$ and $P=2$ for Fig. 17(b). The other parameters of the algorithms are as the same as those in Fig. 16. }
	\label{Fig17}
\end{figure}
According to Remark 5, Fig. 17(a) compares the multiplication complexity of the nonlinear algorithms with respect to the parameter $P$. The proposed Volterra-NKP-NSAF and TFLN-NKP-NSAF algorithms require more multiplicative operations than theircounterpart algorithms. Additionally, the algorithms based on the TFLN structure not only have lower computational costs compared to the Volterra-based approach but also achieve lower steady-state misadjustment. Clearly, from Fig. 17(b), we can draw similar conclusions to those in Fig. 17(a).
\subsection{Active noise control}
In  this section, we introduce the filtered-x NSAF-NKP-II (NKP-FxNSAF), filtered-x RNSAF-NKP-MCC (NKP-FxNSAF-MCC), and filtered-x RNSAF-NKP-LC (NKP-FxNSAF-LC) algorithms. Fig. 18 illustrates the structure of subband NKP decomposition model under ANC scenarios, where $P(z)$ and $S(z)$ denote the primary and secondary paths, respectively, $\hat{S}(z)$ represents the estimated value of $S(z)$.  Compared with the system identification model shown in Fig. 2, the ANC model shown in Fig. 18 only has two differences. Firstly, the input signal of the analysis filters $\{\bm{f}_j\}_{j=1}^N$ is the filtered input signal $x_r^{'}$. Secondly, at the noise-cancelling point, only $e_r$ is a measurable signal, while $d_r$ and $y_r$ are unmeasurable. Therefore, we obtain the subband error signals $\{e_{r,j}\}_{j=1}^N$ by passing the error vector $\bm{e}_r=[e_r, e_{r-1},...,e_{r-L+1}]$ through the analysis filters, i.e.,
\begin{equation}
	\label{078}
\bm{e}_{r,s}\overset{\bigtriangleup}{=}[e_{r,1}, e_{r,2},...,e_{r,N}]=\bm{e}_r^{\text T}\bm{F},
\end{equation}
where $\bm{e}_{r,s}$ denotes the subband error vector. Additionally, the algorithm’s performance indicator average noise reduction (ANR) is defined as \cite{10806852}
\begin{figure}[htp]
	\centering
	\includegraphics[scale=0.33] {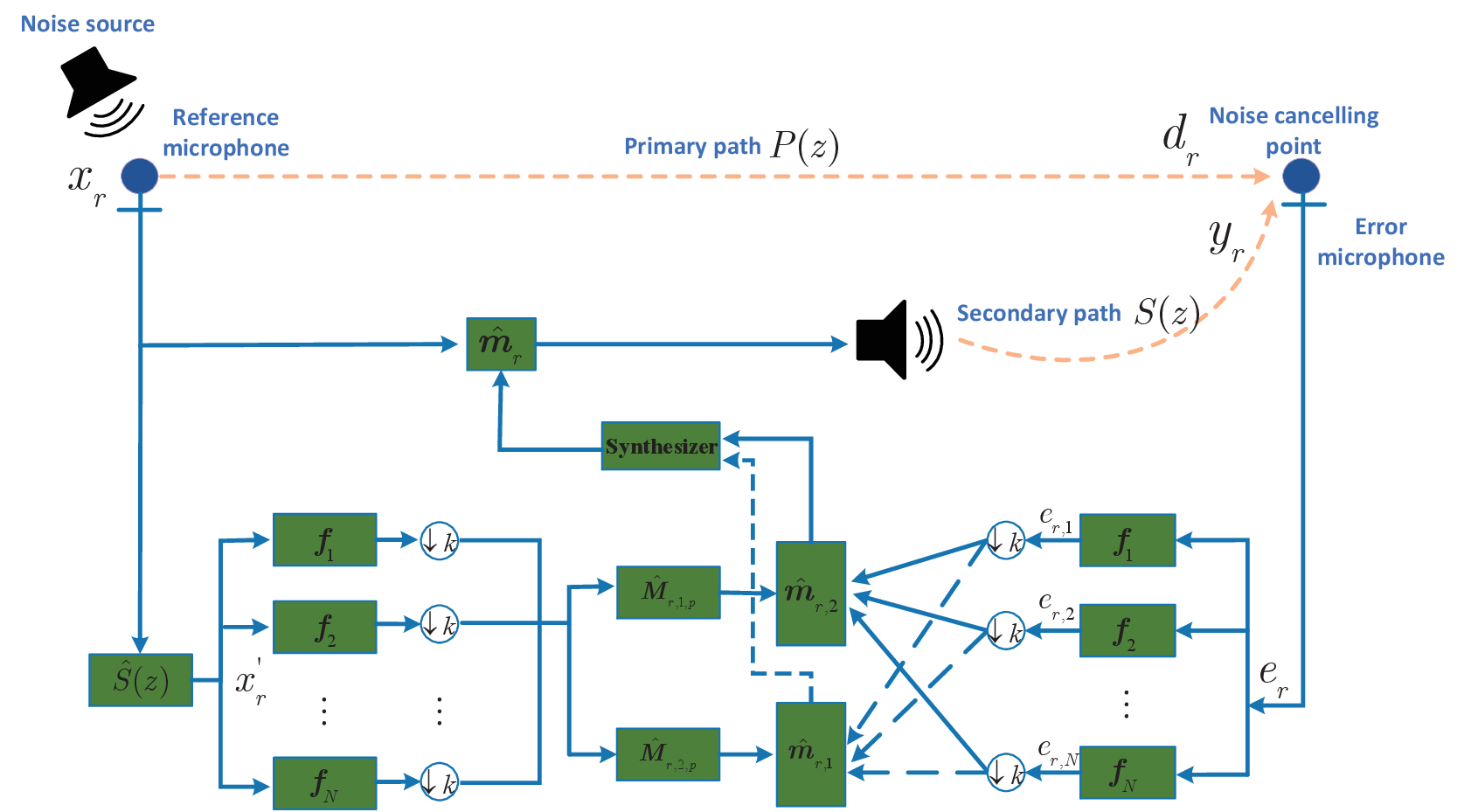}
	\vspace{-1em} \caption{Structure of the subband NKP decomposition model under ANC scenarios.}
	\label{Fig3}
\end{figure}
\begin{figure}[htp]
	\centering  
	\includegraphics[scale=0.29] {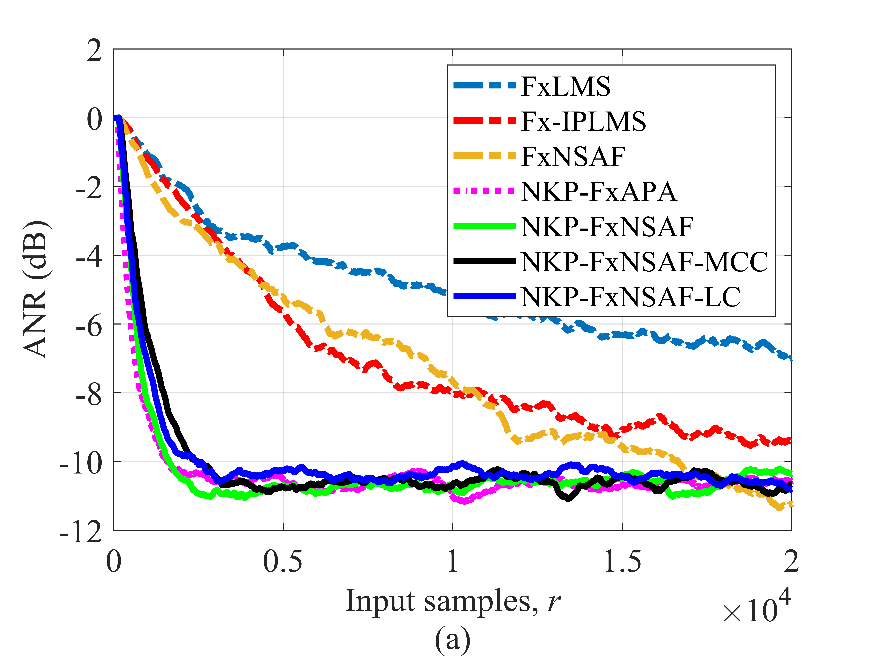}  
	\hspace{0.001ex}
	\includegraphics[scale=0.29] {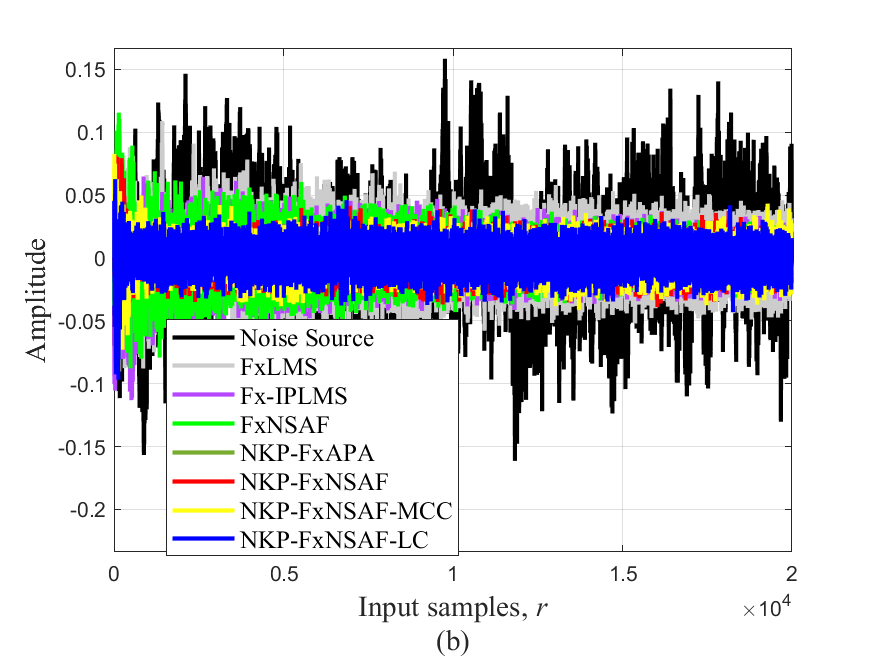} 
	\caption{(a) The ANR curves of several algorithms and (b) the noise reduction results of pink noise. $\mu=0.01$ for the FxLMS algorithm; $\mu=0.1$, $\alpha_1=-0.5$, and $\epsilon_1=10^{-6}$ for the Fx-IPLMS algorithm; $\mu=0.01$ for the FxNSAF algorithm; $C=5$, $\mu_1=\mu_2=0.004$ for the NKP-FxAPA algorithm.}
	\label{Fig18}
\end{figure}

\begin{equation}
	\label{079}
	\text{ANR}(r)\overset{\bigtriangleup}{=}10\text{log}\Big\{\frac{S_e^2(r)}{S_d^2(r)}\Big\},
\end{equation}
where $S_e(r)=\eta S_e(r-1)+(1-\eta)\lvert e_r\lvert$ and $S_d(r)=\eta S_d(r-1)+(1-\eta)\lvert d_r\lvert$ denote the averaged magnitude of residual error and desired signal, and $\eta=0.999$.

Fig. 19(a) compares the ANR curves of the filtered-x least mean square (FxLMS) \cite{elliott2000signal}, filtered-x improved proportionate least mean square (FxIPLMS) \cite{arenas2011combinations}, filtered-x NSAF (FxNSAF), NKP decomposition-based FxAP (NKP-FxAP) \cite{li2025nearest}, filtered-x NSAF-NKP-II (NKP-FxNSAF), filtered-x NSAF-NKP-MCC (NKP-FxNSAF-MCC), and filtered-x NSAF-NKP-LC (NKP-FxNSAF-LC) algorithms. The primary and secondary paths are modeled using finite IR (FIR) filters with transfer functions $P(z)=z^{-3}-0.3z^{-4}+0.2z^{-5}$ and $S(z)=z^{-2}+0.5z^{-5}$ \cite{10806852}, respectively. Clearly, the proposed algorithms exhibit better convergence performance compared to FxLMS, Fx-IPLMS, and FxNSAF algorithms, and simultaneously achieve performance that is consistent with that of the NKP-FxAPA algorithm. As shown in Fig. 19(b), we can obtain the same conclusion as that in Fig. 19(a).
\begin{figure}[htp]
	\centering  
	\includegraphics[scale=0.29] {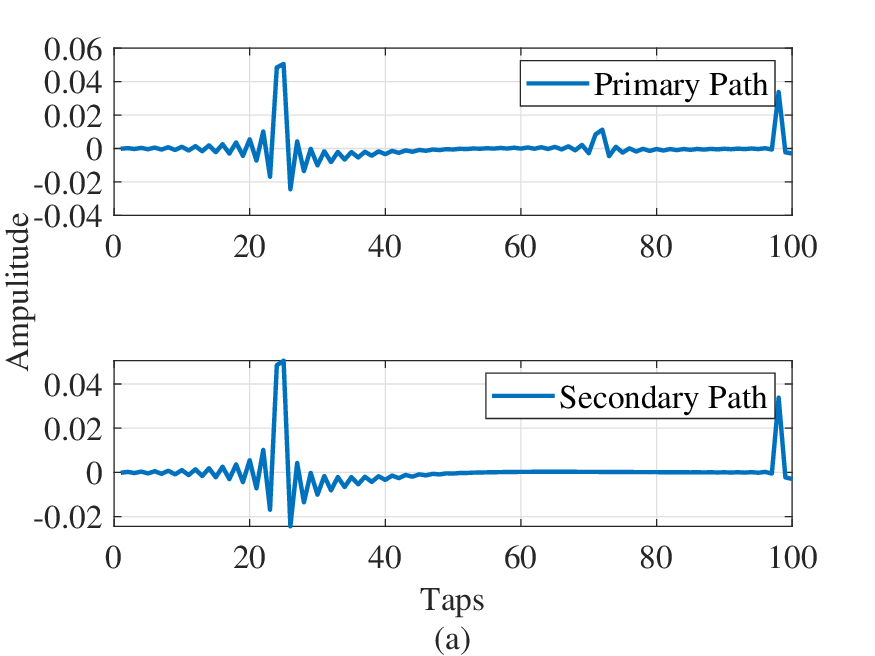} 
	\hspace{0.001ex}
	\includegraphics[scale=0.29] {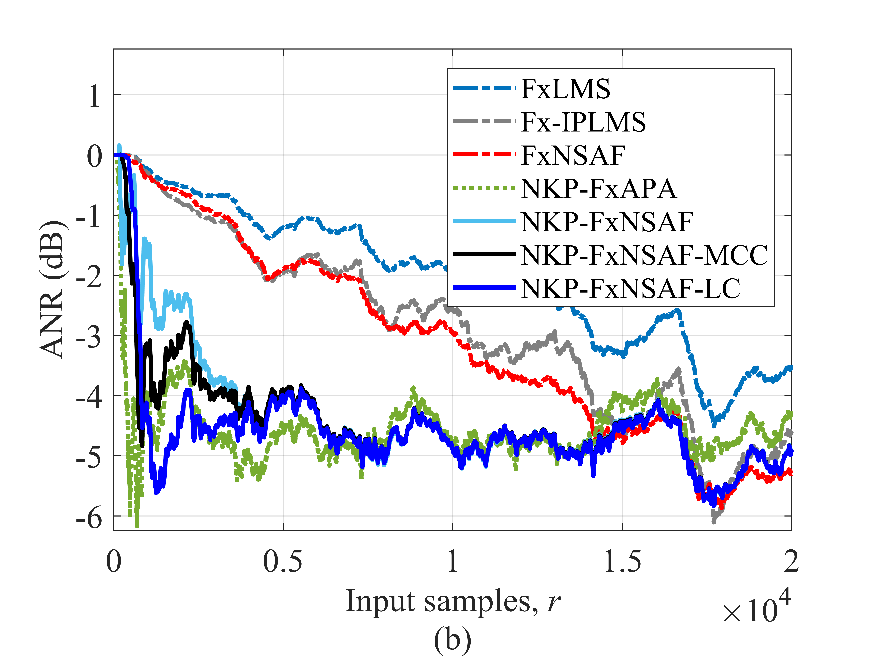}  
	\caption{(a) Time-domain impulse responses of ANC system and (b) the ANR curves of several algorithms. The parameters of the algorithms are as the same as those in Fig. 19. }
	\label{Fig18}
\end{figure}

Fig. 20 (a) depicts the time-domain impulse response of the primary and secondary paths, which were generated using the image method \cite{allen1979image}. The image method has been shown to be effective and accessible for simulating room impulse response and has been widely used in the field of audio signal processing. The noise source signal employed is a real noise recorded from a traction substation \cite{zhu2020robust}. 

Fig. 20 (b) compares the ANR curves of the algorithms in traction substation noise. It is worth noting that we applied the multichannel NKP decomposition-based FxAP algorithm in \cite{li2025nearest} to the single-channel ANC scenario, thereby obtaining the NKP-FxAP algorithm. Clearly, the proposed NKP-FxNSAF, NKP-FxNSAF-MCC, and NKP-FxNSAF-LC algorithms exhibit a faster convergence rate in the ANC system compared to the competing algorithms.  
\section{CONCLUSION}
This paper presents a novel framework that extends the nearest Kronecker product decomposition technique to the subband adaptive filtering domain. The developed NSAF-NKP-I algorithm exhibits accelerated convergence compared to conventional NSAF implementations in sparse system identification tasks but comes at the cost of increased computational complexity. To reduce the computational complexity of the NSAF-NKP-I algorithm, we subsequently developed the NSAF-NKP-II algorithm. For resolving the stability degradation of NSAF-NKP-II under impulsive noise interference, we proposed the RNSAF-NKP-MCC and RNSAF-NKP-LC variants, which demonstrate robust convergence in the impulsive noise scenarios. Furthermore, we compared the computational complexity of the proposed subband NKP-based algorithms with that of existing competing algorithms. To ensure the reliability of the NSAF-NKP-II algorithm in complex nonlinear environments, we further proposed a class of nonlinear subband nearest Kronecker product algorithms, i.e., TFLN-NKP-NSAF and Volterra-NKP-NSAF. Comprehensive simulations spanning echo cancellation, active noise control, system identification, and complex nonlinear scenarios consistently validate the superior performance of the proposed algorithms.

As a potential research direction for the proposed subband NKP-based adaptive filter, adaptively adjusting the tap length of the sub-filters during the transient and steady-state stages could further optimize computational efficiency and convergence performance.

\bibliographystyle{./IEEEtran}
\bibliography{./IEEEabrv,./IEEEexample}

\newpage

\vfill

\end{document}